\documentclass[aps,prl,twocolumn,floatfix,preprintnumbers,superscriptaddress,longbibliography,nofootinbib,10pt]{revtex4-2}

\usepackage[margin=0.7in]{geometry}
\usepackage[utf8]{inputenc}

\usepackage{graphicx}
\usepackage{dcolumn}
\usepackage{bm}

\usepackage{amsmath}
\usepackage{amssymb}
\usepackage{siunitx} 
\usepackage{xcolor} 
\usepackage{ulem}
\usepackage{hyperref}
\usepackage{xcolor}

\begin{document}

\title{
{Archimedean Seesaw: Small Neutrino Masses and Large Lepton-number Violation}
}

\author{Tao Han}
\email{than@pitt.edu}
\affiliation{PITT PACC, Department of Physics and Astronomy,\\ University of Pittsburgh, 3941 O’Hara St., Pittsburgh, PA 15260, USA}

\author{Alejandro Ibarra}
\email{ibarra@tum.de}
\affiliation{Physik-Department, Technische Universit\"at M\"unchen, James-Franck-Stra$\beta$e, 85748, Garching, Germany}

\author{Subhojit Roy}
\email{sroy@anl.gov}
\affiliation{HEP Division, Argonne National Laboratory, 9700 Cass Ave., Argonne, IL 60439, USA}

\author{Martina Sabová}
\email{martina.sabova@tum.de}
\affiliation{Physik-Department, Technische Universit\"at M\"unchen, James-Franck-Stra$\beta$e, 85748, Garching, Germany}

\begin{abstract}
Contrary to the common lore that observable lepton-number violation (LNV) is inevitably suppressed by tiny neutrino masses, we identify a class of seesaw models in which arbitrarily large LNV can naturally coexist with sub-eV neutrino masses. 
We construct a symmetry-protected texture-zero structure in the neutrino Yukawa couplings and heavy Majorana mass matrix that gives rise to the required accidental symmetry, thereby protecting the light neutrinos from acquiring mass even in the presence of arbitrarily large LNV in the heavy sector. 
Small neutrino masses arise naturally from  lifting the texture-zero structure while preserving the underlying symmetry. 
The resulting framework offers a rich and experimentally accessible phenomenology, predicting Heavy Neutral Leptons with sizeable active-sterile mixing over a broad range of experimentally accessible masses, giving rise to observable LNV signatures at collider and intensity-frontier experiments.
\end{abstract}

\preprint{PITT-PACC-2612}

\maketitle

\section{Introduction}
\label{sec:Intro}

The striking differences between the observed neutrino and quark sectors strongly suggest that neutrino masses may not arise solely from the addition of a Dirac mass term to the Lagrangian, but instead could be a hint to the existence of a Majorana mass term, which breaks the total lepton number by two units ($\Delta L = 2$).
With the matter content of the Standard Model (SM), the lowest-dimensional operator violating lepton number appears at dimension-five, the so-called Weinberg operator~\cite{Weinberg:1979sa}
\begin{equation}
    {\kappa \over \Lambda} (L H)(L H),
\end{equation}
where $L$ and $H$ are the SM lepton and Higgs doublets, respectively, $\kappa$ is the Wilson coefficient of the operator and $\Lambda$ denotes the scale of the new physics of underlying lepton-number violation (LNV). As such, any LNV process must then be proportional to the order parameter associated with lepton-number breaking, which in this case is proportional to the neutrino masses. As a consequence, the rates of all low-energy LNV processes are suppressed by the smallness of the observed neutrino masses $m_\nu \sim {\kappa v^2 / 2\Lambda}$, where $v\simeq246~\mathrm{GeV}$ is the Higgs vacuum expectation value, rendering the observation of $\Delta L = 2$ processes extremely challenging experimentally.

On the other hand, the Weinberg operator is expected to be generated by new degrees of freedom beyond the SM, which in turn introduce additional sources of LNV. One of the simplest realizations is the Type-I seesaw mechanism, which introduces fermionic fields that are singlets under the SM gauge group and carry lepton number equal to 1~\cite{Minkowski:1977sc, Gell-Mann:1979vob, Yanagida:1979as,  Mohapatra:1979ia}. These fields are known as right-handed neutrinos, or ``sterile neutrinos''.
Once electroweak (EW) symmetry is spontaneously broken by the vacuum expectation value of the neutral component of the Higgs field, the left- and right-handed neutrinos acquire identical quantum numbers and mix to form mass eigenstates. The LNV associated with the heavy right-handed neutrinos then generates Majorana masses for the light active neutrinos, thus providing an elegant explanation for the smallness of neutrino masses through the breaking of total lepton number. This framework also predicts 
the existence of heavy neutrino mass eigenstates, commonly referred to as Heavy Neutral Leptons (HNLs).

It is of fundamental importance to search for HNLs in high energy experiments as well as in astronomical and cosmological observations. However, in the standard Type-I seesaw, the active-sterile mixing $U_{\ell N}$ and the physical
heavy neutrino mass $M_N$ are related by
\begin{equation}
    U_{\ell N}^2 \, M_N = m_\nu ,
    \label{eq:seesaw-relation}
\end{equation}
so that any LNV rate, controlled by
$U_{\ell N}^2$, is tied directly to the smallness of the observed neutrino mass. For
$M_N$ at the EW scale this forces $U_{\ell N}^2 \lesssim m_\nu/M_N \sim
10^{-13}$, well below the sensitivity of any foreseeable experiment. It is therefore of great significance to explore alternative theories for neutrino mass generation that may deviate from the rigid constraint as in Type-I seesaw of Eq.~(\ref{eq:seesaw-relation}). 
One notable exception is provided by scenarios in which the lepton number is approximately conserved, such as the inverse~\cite{Mohapatra:1986aw,Mohapatra:1986bd} and linear~\cite{Akhmedov:1995ip,Malinsky:2005bi} seesaws. In this case, neutrino masses can be made small by introducing a small lepton-number-breaking parameter, while keeping the active-sterile mixing sizable. However, the approximate conservation of lepton number in the model also suppresses the corresponding LNV signals, making their detection extremely challenging.

In this Letter, we propose a novel scenario and show that the smallness of neutrino masses and the strength of observable LNV need not be controlled by the same parameter as in Eq.~(\ref{eq:seesaw-relation}).
Specifically,
we identify a class of scenarios in which the right-handed neutrinos possess interactions that violate the total lepton number, while the active neutrino masses vanish as a consequence of texture zeros in the Yukawa couplings and the right-handed neutrino mass matrix. We
further show that these textures are associated with an accidental
symmetry of the neutrino mass matrix that protects the zero mode.
Perturbations that lift the texture zeros while preserving the underlying
symmetry of the model then generate naturally small neutrino masses.
In this framework, the resulting HNLs can have sizeable active--sterile
mixing, be copiously produced at colliders, and give rise to observable
LNV signatures while maintaining sub-eV neutrino masses. The underlying mass--mixing relation admits an intuitive
weight--lever-arm interpretation of the seesaw mechanism, motivating us
to dub this scenario the ``Archimedean seesaw''.
Existing
searches already constrain the model parameters, while future experiments
will probe substantial regions of the remaining parameter space.
\section{Accidentally vanishing neutrino masses}
\label{sec:framework}
We consider, for simplicity, a scenario with one lepton doublet, $L=(\nu_L,\ell_L)$, and two right-handed neutrinos,\footnote{Extensions of the model to more leptonic generations, possibly violating lepton flavor, will be discussed in a forthcoming paper.}  $N_{Ri}$, $i=1,2$. The Lagrangian is
\begin{align}
-{\cal L}=y_{i} \overline{L} \widetilde H N_{Ri}+ \frac{1}{2}\overline{N_{Ri}^c} M_{ij} N_{Rj}+{\rm h.c.} \, \, .
\label{eq:Y}
\end{align}
After EW symmetry breaking, the neutrino mass matrix in the basis 
$(\nu_L, N_{Ri}^c)$ is
\begin{align}
\mathcal{M} =
\begin{pmatrix}
0 & m\\
m^T & M
\end{pmatrix} \, \, ,
\label{eq:full-matrix}
\end{align}
with $m_i=y_i v/\sqrt2$.
In the limit $M \gg m$, the effective light neutrino mass is
\begin{align}
m_\nu \simeq - m \, M^{-1} \, m^T \, .
\label{eq:nu-mass}
\end{align}
The requirement that the neutrino mass matrix Eq.~(\ref{eq:nu-mass}) vanishes implies:
\begin{align}
m_1^2 M_{22} - 2m_1 m_2 M_{12}+ m_2^2 M_{11}=0 \, \, .
\end{align}
Notably, the choice $m_1=0$, $M_ {11}=0$ solves this equation. Thus, without loss of generality, we take the matrices as 
\begin{align}
Y=\begin{pmatrix} 0 & y_2 \end{pmatrix} \, \, , ~~~~
M=\begin{pmatrix} 0 & M_{12} \\ M_{12} & M_{22} \end{pmatrix} \, \, , 
\label{eq:example}
\end{align}
that leads to vanishing neutrino masses, while the Lagrangian violates the total lepton number. It is important to stress that if the leading term in the block diagonalization Eq.~(\ref{eq:nu-mass}) vanishes, then the effective light neutrino mass matrix vanishes at all orders.
This can be checked explicitly by replacing the textures Eq.~(\ref{eq:example}) into the ($3\times3$) mass matrix Eq.~(\ref{eq:full-matrix}):
\begin{align}
\mathcal{M}=
\begin{pmatrix}
0 & 0 & m_2\\ 
0 & 0 & M_{12}\\
m_2 & M_{12} & M_{22}
\end{pmatrix}.
\label{eq:3x3_matrix}
\end{align}
The first two rows are linearly dependent, which implies that one eigenvector, which we denote as $\nu$, is massless. On the other hand, the other two eigenvectors, which we denote as $N_1$ and $N_2$, are massive and correspond to the HNLs. 
Assuming for simplicity that all parameters are real, their respective masses are:
\begin{align}
&M_{1,2}=\frac{1}{2}\left(
M_{22}\mp\sqrt{M_{22}^2+4\rho^2}
\right)  \, \, ,
\label{eq:M1M2c}
\end{align}
with $\rho^2=m_2^2+M_{12}^2 = -M_1 M_2$.
The HNLs clearly do not form a Dirac pair, thereby signaling LNV. The mixing matrix that results from the diagonalization ${\cal M}=V D_\nu V^T$, with $D_\nu={\rm diag}(0,M_1,M_2)$ is
\begin{align}
V=\begin{pmatrix}
\dfrac{M_{12}}{\rho}
&
\dfrac{m_2}{\sqrt{\rho^2+M_1^2}}
&
\dfrac{m_2}{\sqrt{\rho^2+M_2^2}}
\\[1em]
-\dfrac{m_2}{\rho}
&
\dfrac{M_{12}}{\sqrt{\rho^2+M_1^2}}
&
\dfrac{M_{12}}{\sqrt{\rho^2+M_2^2}}
\\[1em]
0
&
\dfrac{M_1}{\sqrt{\rho^2+M_1^2}}
&
\dfrac{M_2}{\sqrt{\rho^2+M_2^2}}
\end{pmatrix} \, \, .
\end{align}
Thus,
\begin{equation}
  -V_{\ell N_1}^2 M_1 \;=\; V_{\ell N_2}^2 M_2
  \;=\;
  \frac{m_2^2}{M_2-M_1},
  \label{eq:mixing_mass_signed}
\end{equation}
where $M_1<0$ and $M_2>0$ for any real, nonzero $M_{12}$, $M_{22}$ and $m_2$, as follows from Eq.~\eqref{eq:M1M2c}, since $\rho^2>0$.
In contrast to Eq.~(\ref{eq:seesaw-relation}), the right-hand side is no longer
$m_\nu$, which vanishes identically in this texture-zero limit; instead, it is determined entirely by heavy-sector parameters and can therefore be sizeable. Consequently, the active--sterile mixing is no longer suppressed by the smallness of the neutrino mass, opening the possibility of large LNV signatures.
Eq.~(\ref{eq:mixing_mass_signed}) has the same mathematical structure as the well-known Archimedean balance condition of the lever, with the neutrino masses and active-flavor projections playing the roles of weights and lever arms, respectively.
This correspondence motivates us to refer to this scenario as the texture-zero Archimedean seesaw.

When the texture zeros are lifted by the small mass parameters $\mu_1$ and $\mu_2$
\begin{align}
\mathcal{M}=
\begin{pmatrix}
0 & \mu_1 & m_2\\ 
\mu_1 & \mu_2 & M_{12}\\
m_2 & M_{12} & M_{22}
\end{pmatrix},
\label{eq:3x3_matrix_pert}
\end{align}
one obtains a non-zero active neutrino mass
\begin{align}
\label{mnu}
m_\nu=
-2\mu_1\frac{m_2\,M_{12}}{\rho^2}
+\mu_2\frac{m_2^2}{\rho^2} \, ,
\end{align}
and a non-zero entry in the leptonic mixing matrix
\begin{align}
V_{N_{R2} \nu}\simeq \mu_1\frac{\left(m_2^2-M_{12}^2\right)}{\rho^{3}}+\mu_2\frac{m_2 M_{12}}{\rho^{3}} \, ,
\end{align}
whereas the masses of the HNLs and the rest of the mixing angles are just modified by small corrections ${\cal O}(\mu_1,\mu_2)$. 
If $\mu_1$ and $\mu_2$ are small,  $m_{\nu}$ is naturally
suppressed, while the active--sterile mixing and heavy-sector LNV can
remain sizeable.
An important feature of Eq.~(\ref{mnu}) is that the heavy Majorana mass parameter $M_{22}$, which controls the LNV in the heavy sector, does not appear explicitly in $m_{\nu}$. Consequently, the strength of LNV is no longer tied to the smallness of $m_\nu$, allowing sizeable active--sterile mixing together with potentially observable LNV signatures.
This qualitative decoupling is in sharp contrast to the conventional Type-I seesaw, where the same Majorana mass controls both neutrino masses and LNV rates. 

\begin{figure}[tb]
\label{fig:seesaws}
\centering
\begin{minipage}{0.36\textwidth}
    \centering
    \includegraphics[width=\linewidth]{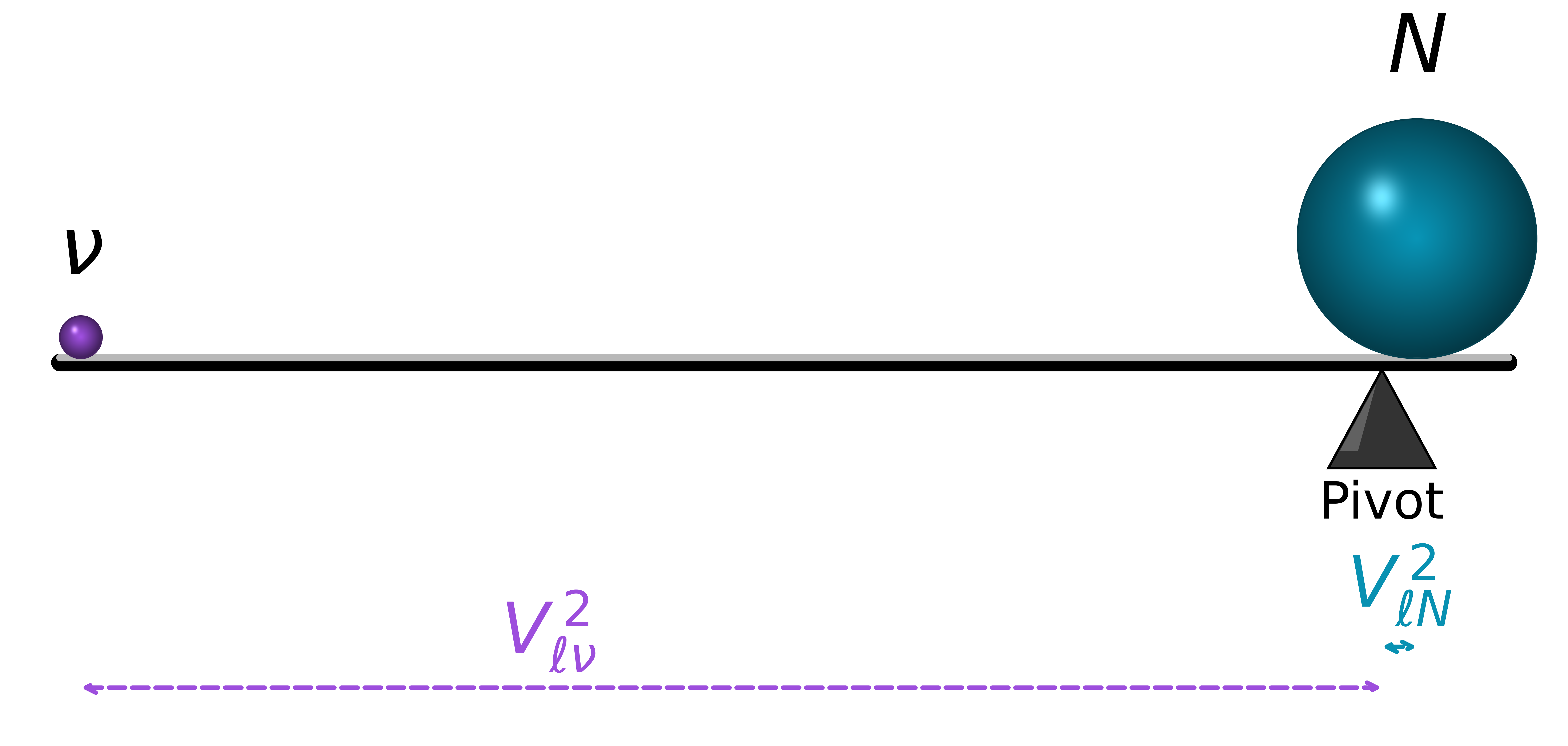}
\end{minipage}
\hfill
\begin{minipage}{0.36\textwidth}
    \centering
    \includegraphics[width=\linewidth]{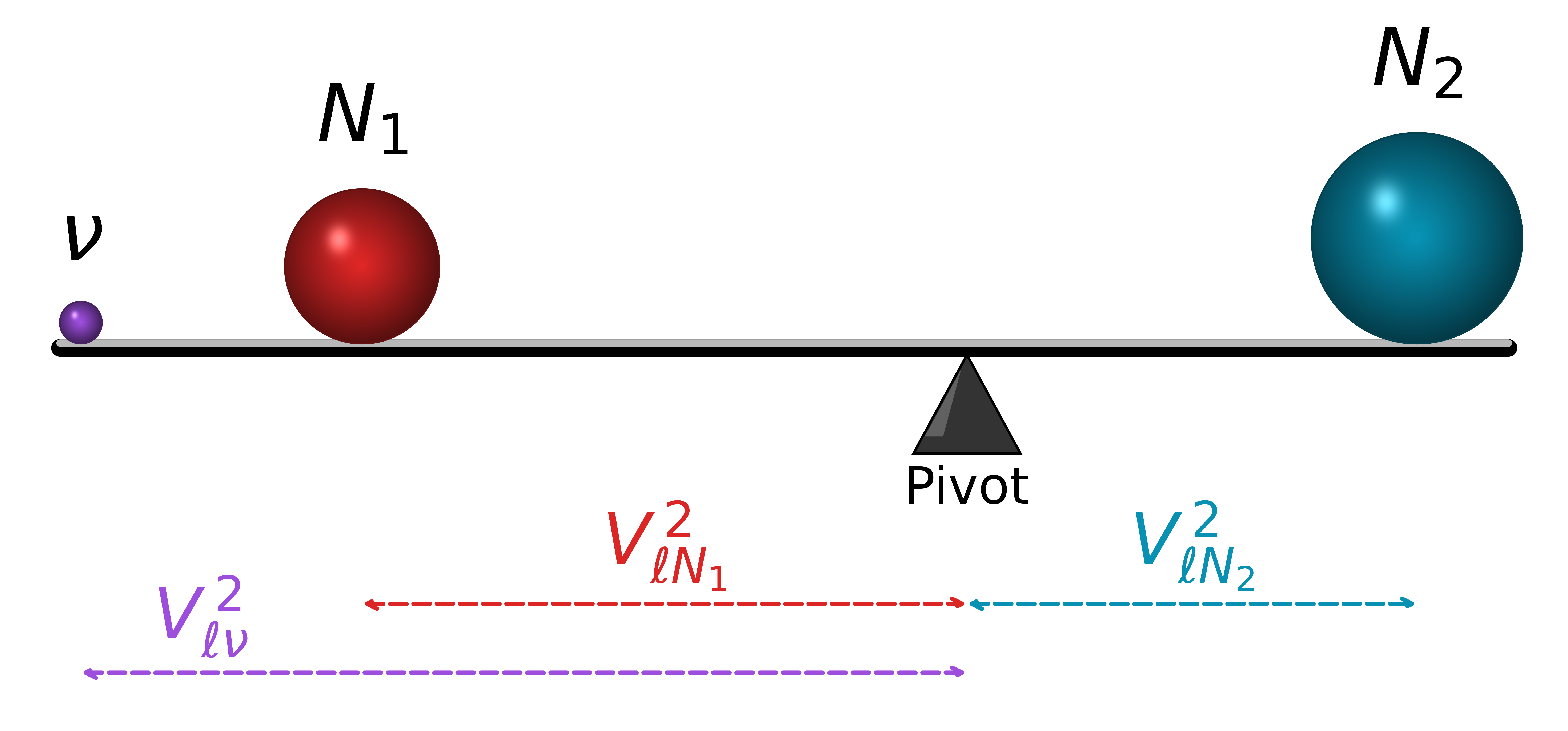}
\end{minipage}
    \caption{Pictorial illustration of the standard Type-I seesaw (top) and the texture-zero Archimedean seesaw (bottom).}
        \label{fig:archimedean}
\end{figure}
Figure \ref{fig:archimedean} provides a geometric interpretation of this distinction.
It exploits the correspondence between the relation $m_\nu V_{\ell\nu}^2 + M_1 V_{\ell N_1}^2 + M_2 V_{\ell N_2}^2 = 0$,
which follows from the condition ${\cal M}_{\ell\ell}=0$ in  Eq.~(\ref{eq:3x3_matrix_pert}). This is directly analogous to the cancellation of the angular momenta induced by the different masses  in a seesaw, corresponding to the  Archimedean
balance condition $\sum_i M_i L_i=0$, with $L_i$ the (signed) distance of the mass $M_i$ from the pivot, representing the corresponding active-flavor projections.
In the standard Type-I seesaw (top panel), following the relation in Eq.~(\ref{eq:seesaw-relation}), the balance of the seesaw with $M_N\gg m_\nu$ requires the heavy mass $M_N$ to be placed very close to the pivot ({\it i.e.} the mixing $V_{\ell N}$ must be tiny). 
This underscores the challenge of realizing a seesaw scenario with only one right-handed neutrino while maintaining testability at collider experiments.
In the presence of two heavy masses, it is possible to place one of them in the seesaw as ``counterweight'' of the other, allowing balance with longer arms  ({\it i.e.}~larger mixing with the active neutrino).
This is the scenario we refer to as the texture-zero Archimedean seesaw (bottom  panel). The balance is generically asymmetric, with $|M_1|\neq M_2$, while $m_\nu$ can still vanish because an accidental symmetry of the texture-zero structure forbids the Weinberg operator, as discussed later, and sizeable LNV can persist. 

In particular, the inverse seesaw can be recovered in the special limit
$M_1\simeq -M_2$, where the two HNLs form an approximate Dirac pair.
In the Archimedean picture, this corresponds to equal masses placed
symmetrically about the pivot. In this limit, approximate lepton-number
conservation protects the light-neutrino mass but also suppresses LNV.
By contrast, in the present texture-zero realization, an accidental
symmetry protects $m_\nu$ without suppressing heavy-sector LNV. This
allows $m_\nu=0$ to coexist with a strongly asymmetric HNL spectrum and
potentially large LNV, motivating the designation of this mechanism as
the texture-zero Archimedean seesaw.
%
%
\section{Texture zeros from an extra $U(1)_X$ symmetry}
\label{sec:model}
The choice Eq.~(\ref{eq:example}) is so far completely phenomenological and imposed {\it ad hoc}. However, 
these structures can arise in models with extra symmetries (different from the total lepton number), thereby naturally justifying the existence of small neutrino masses with HNL interactions that violate the total lepton number. 
We extend the Type-I seesaw model by an additional $U(1)_X$ symmetry and an SM-singlet scalar $\phi$.
The charges of the relevant fields of the model under $U(1)_X$ symmetry are:
\begin{center}
\begin{tabular}{|ccccc|}
\hline 
$L$ & $N_1$ & $N_2$ & $H$ & $\phi$\\
\hline
0 & 1 & 0 & 0& $-1$\\
\hline 
\end{tabular}
\end{center}
The $U(1)_X$ symmetry only allows the following renormalizable terms 
\begin{align}
{\cal L} \supset &y_{2} \bar L \widetilde H N_2 + M_{22} \overline{N_2^c} N_2 + \lambda_\phi \phi \overline{N_1^c} N_2 \, .
\label{eq:dim4}
\end{align}
At dimension five, additional $U(1)_X$-invariant operators may also arise
\begin{align} 
{\cal L} \supset & \alpha\frac{\phi}{\Lambda} \bar L \widetilde H N_1
+ \frac{1}{2}\beta \frac{\phi^2}{\Lambda} \overline{N_1^c} N_1 +{\rm h.c.} \, ,
\label{eq:FN}
\end{align} 
which are suppressed by a heavy scale $\Lambda$. 
When $\phi$~and~$H^0$  acquire vacuum expectation values,  one generates the desired  mass matrix
Eq.~(\ref{eq:3x3_matrix_pert}) with $m_2=y_2 \langle H^0\rangle$,  $M_{12}=\lambda_\phi\langle \phi \rangle$, $\mu_1=\alpha \langle\phi\rangle \langle H^0\rangle/\Lambda$ and $\mu_2=\beta \langle\phi\rangle^2/\Lambda$. Notice that $\mu_1,\mu_2\ll m_2, M_{12},M_{22}$ if  $\Lambda\gg \langle \phi\rangle, \langle H^0\rangle$, thus the active neutrino mass can be made small if $\Lambda$ is sufficiently large.

It is worth noting that in the renormalizable limit, the neutrino mass matrix, Eq.~(\ref{eq:3x3_matrix}), possesses an accidental global symmetry, $U(1)_{\text{acc}}$,  associated with  the exact zero mode  which ensures the vanishing of the light-neutrino mass independently of the heavy-sector parameters. Although preserved by Eq.~(\ref{eq:dim4}), the higher-dimensional operators in Eq.~(\ref{eq:FN}) explicitly break this accidental symmetry, thereby lifting the zero mode and generating the neutrino mass. The explicit construction of the symmetry is presented in the Supplemental Material. In contrast to conventional seesaw scenarios~\cite{ Minkowski:1977sc, Gell-Mann:1979vob, Yanagida:1979as,  Mohapatra:1979ia, Glashow:1979nm,  Gell-Mann:1979vob, Konetschny:1977bn, Magg:1980ut, Lazarides:1980nt, Schechter:1980gr, Cheng:1980qt, Bilenky:1980cx,
 Wyler:1982dd, Mohapatra:1986aw, Mohapatra:1986bd, Akhmedov:1995ip,  Mohapatra:2005wg, Malinsky:2005bi, Shaposhnikov:2006nn,Kersten:2007vk,Gavela:2009cd,Ibarra:2010xw,Antusch:2015mia, Moffat:2017feq},
 the interplay between the texture-zero structure, the accidental
$U(1)_{\rm acc}$ symmetry, and its explicit breaking by higher-dimensional
operators defines a distinct class of low-scale seesaw mechanisms.
Furthermore, this~$U(1)_{\rm acc}$~symmetry assures the zero neutrino mass technically natural and the non-zero mass parametrically small, from spontaneously symmetry breaking and  higher dimensional~operators.
%
%
%
\section{Phenomenological Signatures and Experimental Sensitivity}
The Archimedean seesaw scenario predicts sizable active--sterile  mixing
maintaining naturally small neutrino masses, thereby opening the possibility of
observable LNV signatures over a broad region of
parameter space. Upon diagonalizing Eq.~(\ref{eq:3x3_matrix}), the charged-current
interaction between the HNLs and the charged lepton
$\ell$~is
\begin{align}
\label{eq:CC}
{\cal  L}_{\rm CC} & \supset -\frac{g }{\sqrt{2}}\,W^-_\mu\,\overline{\ell}\gamma^\mu \left(V_{\ell N_1}\ N_{1}+V_{\ell N_2}\ N_{2} \right) +\text{h.c.} \, ,
\end{align} 
where the mixing angles $V_{\ell N_1}$ and $V_{\ell N_2}$ are fixed
entirely by the heavy-sector parameters through
Eq.~(\ref{eq:mixing_mass_signed}).
In the hierarchical limit $M_2\gg |M_1|$, one obtains
\begin{equation}
V_{\ell N_1}^2 |M_1| M_2 \sim  m_2^2 = {1\over 2}y_2^2v^2\ ,
\label{eq:m1m2}
\end{equation}
which directly correlates the observable HNL mass and active--sterile mixing with the underlying heavy-sector scale. For example, for suitable choices of the Yukawa coupling, {\it e.g.} $y_2$ $\sim$ $10^{-2}\ (10^{-5})$, corresponding to $m_2\simeq1~{\rm GeV}\ (1~{\rm MeV})$, the Archimedean seesaw can naturally admit both Majorana neutrinos within experimentally accessible mass ranges, opening the possibility of complementary probes of the light and heavy HNL states at intensity-frontier and collider experiments.

\begin{figure}[tb]
\includegraphics[scale=0.7]{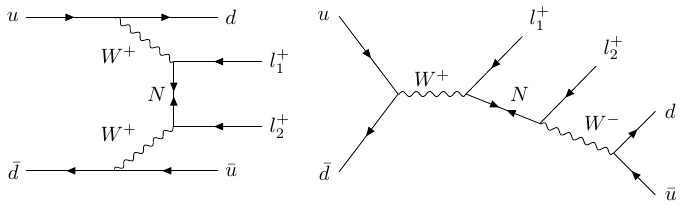}
    \caption{Representative Feynman diagrams for the LNV process mediated by a heavy Majorana neutrino $N$. Left: (a) same-sign vector-boson-fusion topology. Right: (b) resonant Drell--Yan production channel.}
    \label{fig:feyn}
\end{figure}

The Majorana nature of the neutrino gives rise to $\Delta L=2$ processes. The most-thought-after process is the neutrinoless double-beta decay \cite{Blennow:2010th, Mitra:2011qr, Atre:2009rg}, as shown by the representative Feynman diagram in Fig.~\ref{fig:feyn}(a). 
The matrix element scales schematically as
\begin{equation}
{\cal A}\propto\sum_i V_{\ell_1 N_i}
\frac{M_{N_i}}{p^2-M_{N_i}^2} V_{\ell_2 N_i} .
\nonumber
\end{equation}
For a neutrino with mass lighter than the typical momentum transfer $p^2$, such as active neutrinos in the SM, 
one probes the effective mass 
${\cal A}\propto V_{e\nu_i}^2 M_{\nu_i}$. 
For heavier HNLs $M_{N_i}^2\gg p^2$, it is sensitive to  
${\cal A}\propto V_{eN_i}^2/M_{N_i}$. 
We write the cross section for like-sign dilepton production with arbitrary flavor combinations
\begin{equation}
\sigma(W^\pm W^\pm\rightarrow\ell_1^\pm\ell_2^\pm)
=
(2-\delta_{\ell_1\ell_2})
|V_{\ell_1N}V_{\ell_2N}|^2
\sigma_0(WW),
\end{equation}
where $\sigma_0(WW)$ denotes the bare EW production cross section~\cite{Atre:2009rg}.
At the $14$ TeV LHC,
$\sigma_0$ reaches approximately $80$ pb for
$M_N\simeq200$ GeV and decreases to
$\sim30$~fb at $M_N\simeq1$ TeV,
potentially yielding $|V_{\ell_1 N} V_{\ell_2 N}|^2 \times 10^5$ like-sign dilepton events per ab$^{-1}$.
To facilitate comparison with conventional HNL searches, here and henceforth, we use
$M_N\equiv |M_1|$ and $V_{\ell N}\equiv V_{\ell N_1}$.
We note that when $N_1$ and $N_2$ become nearly degenerate, corresponding to the $M_{22}\to0$ (inverse-seesaw) limit, destructive interference between their amplitudes, as implied by Eq.~(\ref{eq:CC}), partially suppresses the LNV rate.

\begin{figure*}[tb]
    \centering
\includegraphics[width=0.41\linewidth]{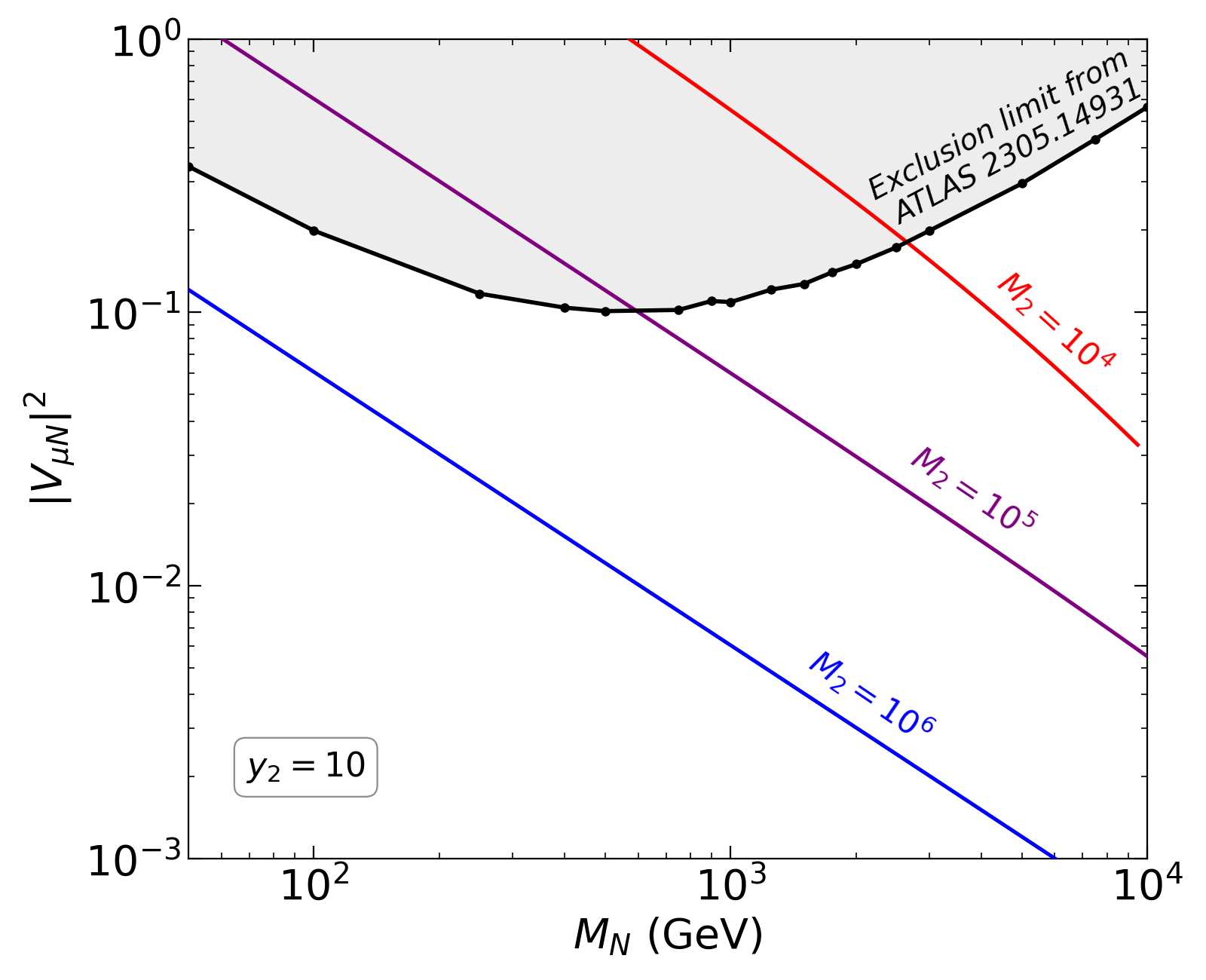}~~~~~~~~~~~~~~~~~
\includegraphics[width=0.41\linewidth]{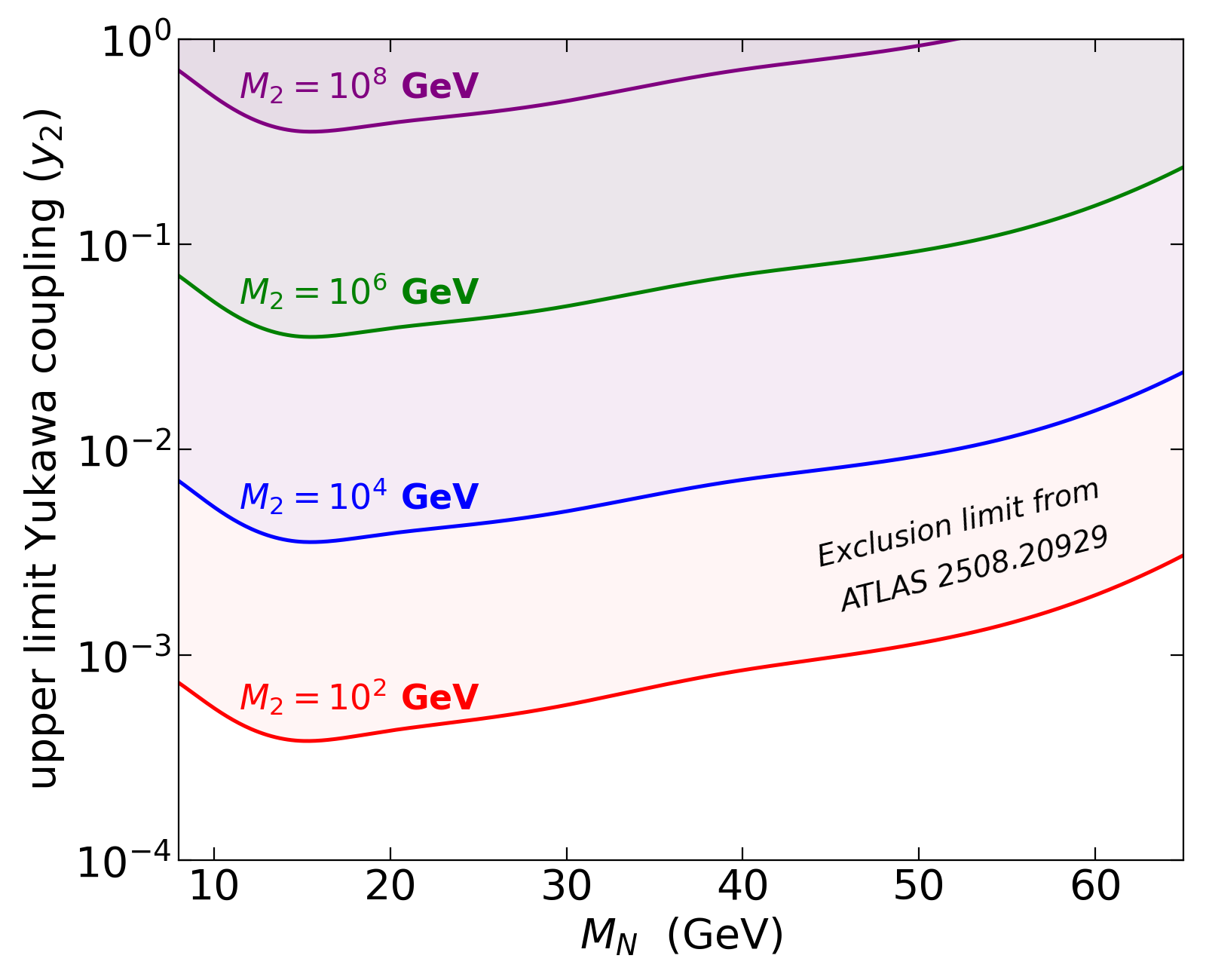}
    \caption{Left: ATLAS exclusion limit in ($M_N-|V_{\mu N}|^2$) plane from the same-sign vector boson scattering~\cite{ATLAS:2023tkz,ATLAS:2024rzi} process. 
    To adopt the conventional HNL notation, we use
$M_N \equiv |M_1|$ and
$V_{\mu N} \equiv V_{\mu N_1}$.
    The colored diagonal lines present the heavy-sector mass scale $M_2$ for $y_2=10$. Right: Corresponding recast of the latest ATLAS prompt resonant HNL search~\cite{ATLAS:2025lva} in the $(M_N-y_2)$ plane. The colored curves represent constant $M_2$, with the parameter space above (below) the curves excluded (allowed).}
    \label{fig:Const}
\end{figure*}

The recent ATLAS searches for Majorana neutrinos in same-sign vector boson scattering probe HNL mass up to $\sim20$ TeV~\cite{ATLAS:2023tkz,ATLAS:2024rzi}. However, the corresponding exclusion, shown in the left panel of Fig.~\ref{fig:Const}, is less stringent than even the bounds from EW precision observables~\cite{delAguila:2008pw,Akhmedov:2013hec,deBlas:2013gla,Basso:2013jka,Antusch:2014woa,Antusch:2015mia,Chrzaszcz:2019inj,Bryman:2021teu,Blennow:2023mqx}, leaving most of the parameter space with sizeable Yukawa couplings and larger $M_2$ unconstrained, as depicted by the straight lines. Future measurements at the HL-LHC are expected to substantially improve the sensitivity to this channel, extending the direct reach into the unexplored region of the model parameter space.

Even more promising is resonant HNL production \cite{Keung:1983uu} $q\bar q\rightarrow W^\pm\rightarrow\ell^\pm N_i$,
followed by the Majorana decay
$N_i\rightarrow\ell^\pm jj$,
which gives rise to the characteristic same-sign dilepton signature
$\ell^\pm\ell^\pm jj$, as depicted in Fig.~\ref{fig:feyn}(b).
The corresponding cross section can be approximated by
\begin{equation}
\sigma(pp\rightarrow\ell^\pm\ell^\pm jj)
\simeq
V_{\ell N_1}^2\sigma_0(M_1)
+
V_{\ell N_2}^2\sigma_0(M_2),
\end{equation}
where $\sigma_0(M)$ depends only on the HNL mass~\cite{Atre:2009rg}.
For $M_N<M_W$, the Drell--Yan production cross section may reach
$\mathcal O(50~{\rm nb})$,
while for EW-scale HNLs it decreases from
$\sim500$~fb at
$M_N\simeq200$~GeV
to
$\sim1$~fb
at
$M_N\simeq1$~TeV.
The latest ATLAS search~\cite{ATLAS:2025lva} for this channel sets the  limit $|V_{\mu N}|^2\lesssim3\times10^{-6}$ for $M_N=10$--$65$ GeV. Recasting this bound in the $(M_N-y_2)$ plane, as shown in the right panel of Fig.~\ref{fig:Const}, demonstrates that the current sensitivity already excludes heavy-sector scales of $M_2\sim10^8$ GeV for $y_2\gtrsim1$, or, $M_2\sim100$ GeV even for $y_2\gtrsim10^{-3}$.~Future LHC searches will further extend the sensitivity toward smaller $y_2/M_2$, probing a substantial fraction of the currently unexplored~parameter~space.

Complementary constraints on HNLs arise from rare meson and tau decays, EW precision measurements, beam-dump experiments, cosmology, and astrophysical observations. These constraints, along with projected sensitivities of future facilities, are summarized in Fig.~\ref{fig:Const2}. More details are provided in the Supplemental Material and references therein.

Unlike the conventional Type-I seesaw, described by Eq.~(\ref{eq:seesaw-relation}) (black dashed line in Fig.~\ref{fig:Const2}), the texture-zero Archimedean seesaw predicts a family of experimentally testable constant-$M_2$ trajectories, shown by the solid colored lines for $M_2 = 10^4, 10^6,1 0^8, 10^{10}$, and $10^{12}$ GeV with the benchmark choice $y_2=0.1$. These trajectories provide a direct mapping between the observable $M_N-|V_{\mu N}|^2$ plane and the underlying heavy-sector scale. Their intersections with the current exclusion limits delineate the portions of parameter space already ruled out, whereas the projected sensitivities indicate the regions that will be explored by future experiments. Consequently, every new exclusion or discovery in the $M_N-|V_{\mu N}|^2$ plane translates directly into a constraint or determination of $M_2$, making the Archimedean seesaw a highly predictive and experimentally testable framework.
The mass range, 1~MeV~$< M_N <$~10~TeV, shown in Fig.~\ref{fig:Const2} represents only a subset of the full parameter space; the seesaw scenario  is not limited to this interval and can be readily
extended to both lighter and heavier~$M_N$.

While the present work has been formulated for one active flavor, its extension to a realistic three-generation framework naturally introduces $CP$-violation and flavor-dependent active--sterile mixings, opening the possibility of low-scale leptogenesis as well as charged-lepton-flavor-violating processes, such as $\mu\to e\gamma$, $\mu\to3e$,~$\mu$--$e$ conversion, and rare $\tau$ decays~\cite{Ilakovac:1994kj,Alonso:2012ji}. Together, these prospects highlight the texture-zero~Archimedean~seesaw~as a versatile framework whose phenomenological implications extend well beyond the signatures explored in this work~\footnote{A detailed study of these possibilities is left for future work.}. 
%
%
\section{Summary and Conclusions}
We have shown that the commonly assumed connection between tiny neutrino masses and suppressed LNV is not an intrinsic consequence of the seesaw mechanism, but rather arises only in a restricted class of realizations. In contrast,  we identified a broad class of neutrino-mass structures in which an enhanced accidental symmetry protects the light-neutrino masses while allowing arbitrarily large LNV in the heavy Majorana sector. The accidental symmetry constitutes the essential ingredient of this mechanism and is independent of its specific ultraviolet realization, thus defining a generic class of low-scale seesaw scenarios. As an explicit realization, we constructed a symmetry-protected texture-zero framework in which the accidental symmetry emerges naturally, and higher-dimensional operators lift the protected limit to generate naturally small neutrino masses. Consequently, naturally small neutrino masses, sizeable active-sterile mixing, and observable HNLs can coexist over a broad parameter space. More broadly, our results demonstrate that the mechanism protecting light-neutrino masses need not be the same as the one governing observable LNV, thereby opening a qualitatively new direction for constructing experimentally testable seesaw theories.
\begin{figure*}[tb]
\includegraphics[width=17.5cm,height=7.25cm]{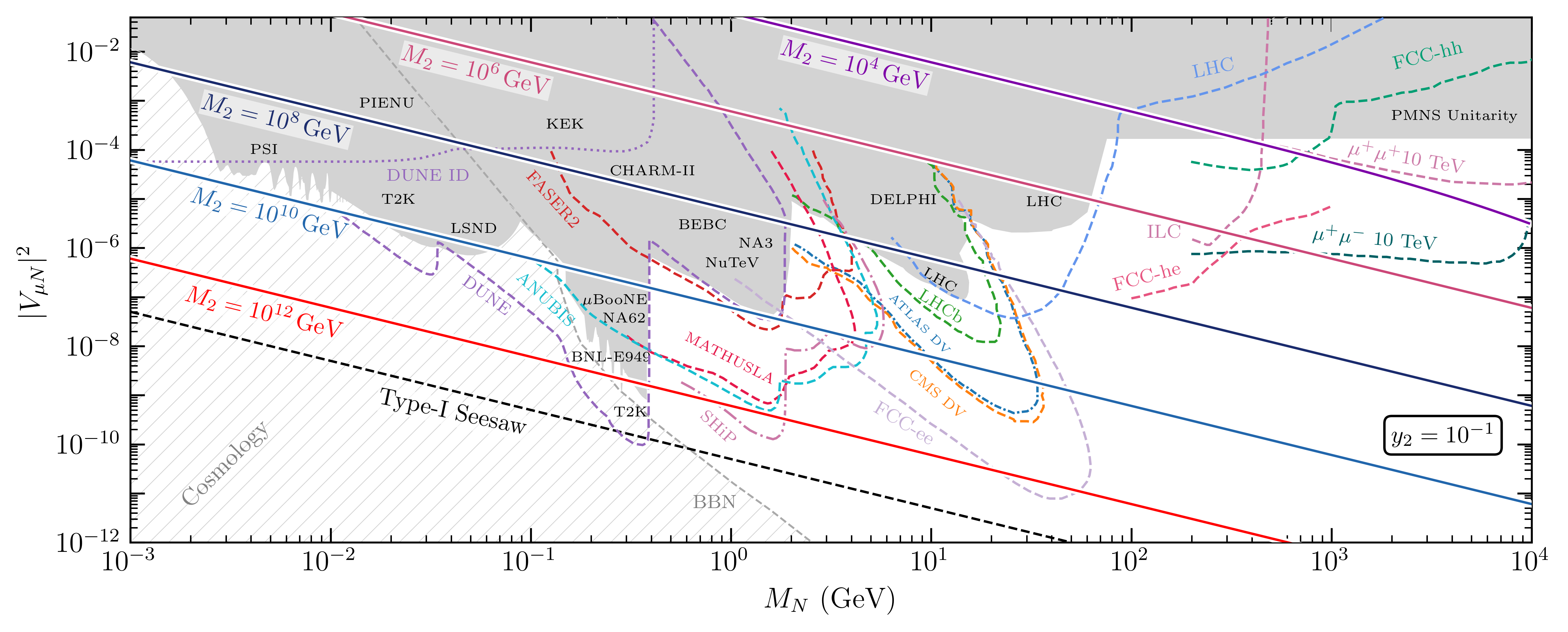}
    \caption{Current experimental constraints (gray shaded regions) and projected sensitivities of future experiments (colored dashed contours) in the $(M_N,|V_{\mu N}|^2)$ plane, assuming muon-flavor dominance. Here, $M_N \equiv |M_1|$ and
$V_{\mu N} \equiv V_{\mu N_1}$,
following the conventional HNL notation. The colored diagonal lines correspond to constant values of the heavy-sector mass parameter $M_2$ for the benchmark $y_2=0.1$. The black dashed line denotes the conventional Type-I seesaw relation. Smaller Yukawa couplings shift the constant-$M_2$ contours downward according to $|V_{\mu N}|^2\propto y_2^2$.}
    \label{fig:Const2}
\end{figure*}
\section*{Acknowledgments}
\noindent
TH was supported in part by the U.S.~Department of Energy under grant No.~DE-SC0007914 and in  part by Pitt PACC. He would also like to thank the Aspen Center for Physics for hospitality during the final stage of this project, which is supported by NSF Grant PHY-2210452. AI and MS were supported by the Collaborative Research Center SFB1258 and by the Deutsche Forschungsgemeinschaft (DFG, German Research Foundation) under Germany's Excellence 
Strategy-EXC-2094-390783311. SR is supported by the U.S.~Department of Energy under contracts No.~DEAC02-06CH11357 at the Argonne National Laboratory. SR would like to thank the University of Chicago, Fermilab, Perimeter Institute, Aspen Center for Physics and Pitt PACC at the University of Pittsburgh, where a significant part of this work was carried out.
\bibliography{references}

\begin{thebibliography}{92}%
\makeatletter
\providecommand \@ifxundefined [1]{%
 \@ifx{#1\undefined}
}%
\providecommand \@ifnum [1]{%
 \ifnum #1\expandafter \@firstoftwo
 \else \expandafter \@secondoftwo
 \fi
}%
\providecommand \@ifx [1]{%
 \ifx #1\expandafter \@firstoftwo
 \else \expandafter \@secondoftwo
 \fi
}%
\providecommand \natexlab [1]{#1}%
\providecommand \enquote  [1]{``#1''}%
\providecommand \bibnamefont  [1]{#1}%
\providecommand \bibfnamefont [1]{#1}%
\providecommand \citenamefont [1]{#1}%
\providecommand \href@noop [0]{\@secondoftwo}%
\providecommand \href [0]{\begingroup \@sanitize@url \@href}%
\providecommand \@href[1]{\@@startlink{#1}\@@href}%
\providecommand \@@href[1]{\endgroup#1\@@endlink}%
\providecommand \@sanitize@url [0]{\catcode `\\12\catcode `\$12\catcode `\&12\catcode `\#12\catcode `\^12\catcode `\_12\catcode `\%12\relax}%
\providecommand \@@startlink[1]{}%
\providecommand \@@endlink[0]{}%
\providecommand \url  [0]{\begingroup\@sanitize@url \@url }%
\providecommand \@url [1]{\endgroup\@href {#1}{\urlprefix }}%
\providecommand \urlprefix  [0]{URL }%
\providecommand \Eprint [0]{\href }%
\providecommand \doibase [0]{https://doi.org/}%
\providecommand \selectlanguage [0]{\@gobble}%
\providecommand \bibinfo  [0]{\@secondoftwo}%
\providecommand \bibfield  [0]{\@secondoftwo}%
\providecommand \translation [1]{[#1]}%
\providecommand \BibitemOpen [0]{}%
\providecommand \bibitemStop [0]{}%
\providecommand \bibitemNoStop [0]{.\EOS\space}%
\providecommand \EOS [0]{\spacefactor3000\relax}%
\providecommand \BibitemShut  [1]{\csname bibitem#1\endcsname}%
\let\auto@bib@innerbib\@empty
\bibitem [{\citenamefont {Weinberg}(1979)}]{Weinberg:1979sa}%
  \BibitemOpen
  \bibfield  {author} {\bibinfo {author} {\bibfnamefont {S.}~\bibnamefont {Weinberg}},\ }\bibfield  {title} {\bibinfo {title} {{Baryon and Lepton Nonconserving Processes}},\ }\href {https://doi.org/10.1103/PhysRevLett.43.1566} {\bibfield  {journal} {\bibinfo  {journal} {Phys. Rev. Lett.}\ }\textbf {\bibinfo {volume} {43}},\ \bibinfo {pages} {1566} (\bibinfo {year} {1979})}\BibitemShut {NoStop}%
\bibitem [{\citenamefont {Minkowski}(1977)}]{Minkowski:1977sc}%
  \BibitemOpen
  \bibfield  {author} {\bibinfo {author} {\bibfnamefont {P.}~\bibnamefont {Minkowski}},\ }\bibfield  {title} {\bibinfo {title} {{$\mu \to e\gamma$ at a Rate of One Out of $10^{9}$ Muon Decays?}},\ }\href {https://doi.org/10.1016/0370-2693(77)90435-X} {\bibfield  {journal} {\bibinfo  {journal} {Phys. Lett. B}\ }\textbf {\bibinfo {volume} {67}},\ \bibinfo {pages} {421} (\bibinfo {year} {1977})}\BibitemShut {NoStop}%
\bibitem [{\citenamefont {Gell-Mann}\ \emph {et~al.}(1979)\citenamefont {Gell-Mann}, \citenamefont {Ramond},\ and\ \citenamefont {Slansky}}]{Gell-Mann:1979vob}%
  \BibitemOpen
  \bibfield  {author} {\bibinfo {author} {\bibfnamefont {M.}~\bibnamefont {Gell-Mann}}, \bibinfo {author} {\bibfnamefont {P.}~\bibnamefont {Ramond}},\ and\ \bibinfo {author} {\bibfnamefont {R.}~\bibnamefont {Slansky}},\ }\bibfield  {title} {\bibinfo {title} {{Complex Spinors and Unified Theories}},\ }\href@noop {} {\bibfield  {journal} {\bibinfo  {journal} {Conf. Proc. C}\ }\textbf {\bibinfo {volume} {790927}},\ \bibinfo {pages} {315} (\bibinfo {year} {1979})},\ \Eprint {https://arxiv.org/abs/1306.4669} {arXiv:1306.4669 [hep-th]} \BibitemShut {NoStop}%
\bibitem [{\citenamefont {Yanagida}(1979)}]{Yanagida:1979as}%
  \BibitemOpen
  \bibfield  {author} {\bibinfo {author} {\bibfnamefont {T.}~\bibnamefont {Yanagida}},\ }\bibfield  {title} {\bibinfo {title} {{Horizontal gauge symmetry and masses of neutrinos}},\ }\href@noop {} {\bibfield  {journal} {\bibinfo  {journal} {Conf. Proc. C}\ }\textbf {\bibinfo {volume} {7902131}},\ \bibinfo {pages} {95} (\bibinfo {year} {1979})}\BibitemShut {NoStop}%
\bibitem [{\citenamefont {Mohapatra}\ and\ \citenamefont {Senjanovic}(1980)}]{Mohapatra:1979ia}%
  \BibitemOpen
  \bibfield  {author} {\bibinfo {author} {\bibfnamefont {R.~N.}\ \bibnamefont {Mohapatra}}\ and\ \bibinfo {author} {\bibfnamefont {G.}~\bibnamefont {Senjanovic}},\ }\bibfield  {title} {\bibinfo {title} {{Neutrino Mass and Spontaneous Parity Nonconservation}},\ }\href {https://doi.org/10.1103/PhysRevLett.44.912} {\bibfield  {journal} {\bibinfo  {journal} {Phys. Rev. Lett.}\ }\textbf {\bibinfo {volume} {44}},\ \bibinfo {pages} {912} (\bibinfo {year} {1980})}\BibitemShut {NoStop}%
\bibitem [{\citenamefont {Mohapatra}(1986)}]{Mohapatra:1986aw}%
  \BibitemOpen
  \bibfield  {author} {\bibinfo {author} {\bibfnamefont {R.~N.}\ \bibnamefont {Mohapatra}},\ }\bibfield  {title} {\bibinfo {title} {{Mechanism for Understanding Small Neutrino Mass in Superstring Theories}},\ }\href {https://doi.org/10.1103/PhysRevLett.56.561} {\bibfield  {journal} {\bibinfo  {journal} {Phys. Rev. Lett.}\ }\textbf {\bibinfo {volume} {56}},\ \bibinfo {pages} {561} (\bibinfo {year} {1986})}\BibitemShut {NoStop}%
\bibitem [{\citenamefont {Mohapatra}\ and\ \citenamefont {Valle}(1986)}]{Mohapatra:1986bd}%
  \BibitemOpen
  \bibfield  {author} {\bibinfo {author} {\bibfnamefont {R.~N.}\ \bibnamefont {Mohapatra}}\ and\ \bibinfo {author} {\bibfnamefont {J.~W.~F.}\ \bibnamefont {Valle}},\ }\bibfield  {title} {\bibinfo {title} {{Neutrino Mass and Baryon Number Nonconservation in Superstring Models}},\ }\href {https://doi.org/10.1103/PhysRevD.34.1642} {\bibfield  {journal} {\bibinfo  {journal} {Phys. Rev. D}\ }\textbf {\bibinfo {volume} {34}},\ \bibinfo {pages} {1642} (\bibinfo {year} {1986})}\BibitemShut {NoStop}%
\bibitem [{\citenamefont {Akhmedov}\ \emph {et~al.}(1996)\citenamefont {Akhmedov}, \citenamefont {Lindner}, \citenamefont {Schnapka},\ and\ \citenamefont {Valle}}]{Akhmedov:1995ip}%
  \BibitemOpen
  \bibfield  {author} {\bibinfo {author} {\bibfnamefont {E.~K.}\ \bibnamefont {Akhmedov}}, \bibinfo {author} {\bibfnamefont {M.}~\bibnamefont {Lindner}}, \bibinfo {author} {\bibfnamefont {E.}~\bibnamefont {Schnapka}},\ and\ \bibinfo {author} {\bibfnamefont {J.~W.~F.}\ \bibnamefont {Valle}},\ }\bibfield  {title} {\bibinfo {title} {{Left-right symmetry breaking in NJL approach}},\ }\href {https://doi.org/10.1016/0370-2693(95)01504-3} {\bibfield  {journal} {\bibinfo  {journal} {Phys. Lett. B}\ }\textbf {\bibinfo {volume} {368}},\ \bibinfo {pages} {270} (\bibinfo {year} {1996})},\ \Eprint {https://arxiv.org/abs/hep-ph/9507275} {arXiv:hep-ph/9507275} \BibitemShut {NoStop}%
\bibitem [{\citenamefont {Malinsky}\ \emph {et~al.}(2005)\citenamefont {Malinsky}, \citenamefont {Romao},\ and\ \citenamefont {Valle}}]{Malinsky:2005bi}%
  \BibitemOpen
  \bibfield  {author} {\bibinfo {author} {\bibfnamefont {M.}~\bibnamefont {Malinsky}}, \bibinfo {author} {\bibfnamefont {J.~C.}\ \bibnamefont {Romao}},\ and\ \bibinfo {author} {\bibfnamefont {J.~W.~F.}\ \bibnamefont {Valle}},\ }\bibfield  {title} {\bibinfo {title} {{Novel supersymmetric SO(10) seesaw mechanism}},\ }\href {https://doi.org/10.1103/PhysRevLett.95.161801} {\bibfield  {journal} {\bibinfo  {journal} {Phys. Rev. Lett.}\ }\textbf {\bibinfo {volume} {95}},\ \bibinfo {pages} {161801} (\bibinfo {year} {2005})},\ \Eprint {https://arxiv.org/abs/hep-ph/0506296} {arXiv:hep-ph/0506296} \BibitemShut {NoStop}%
\bibitem [{\citenamefont {Glashow}(1980)}]{Glashow:1979nm}%
  \BibitemOpen
  \bibfield  {author} {\bibinfo {author} {\bibfnamefont {S.~L.}\ \bibnamefont {Glashow}},\ }\bibfield  {title} {\bibinfo {title} {{The Future of Elementary Particle Physics}},\ }\href {https://doi.org/10.1007/978-1-4684-7197-7_15} {\bibfield  {journal} {\bibinfo  {journal} {NATO Sci. Ser. B}\ }\textbf {\bibinfo {volume} {61}},\ \bibinfo {pages} {687} (\bibinfo {year} {1980})}\BibitemShut {NoStop}%
\bibitem [{\citenamefont {Konetschny}\ and\ \citenamefont {Kummer}(1977)}]{Konetschny:1977bn}%
  \BibitemOpen
  \bibfield  {author} {\bibinfo {author} {\bibfnamefont {W.}~\bibnamefont {Konetschny}}\ and\ \bibinfo {author} {\bibfnamefont {W.}~\bibnamefont {Kummer}},\ }\bibfield  {title} {\bibinfo {title} {{Nonconservation of Total Lepton Number with Scalar Bosons}},\ }\href {https://doi.org/10.1016/0370-2693(77)90407-5} {\bibfield  {journal} {\bibinfo  {journal} {Phys. Lett. B}\ }\textbf {\bibinfo {volume} {70}},\ \bibinfo {pages} {433} (\bibinfo {year} {1977})}\BibitemShut {NoStop}%
\bibitem [{\citenamefont {Magg}\ and\ \citenamefont {Wetterich}(1980)}]{Magg:1980ut}%
  \BibitemOpen
  \bibfield  {author} {\bibinfo {author} {\bibfnamefont {M.}~\bibnamefont {Magg}}\ and\ \bibinfo {author} {\bibfnamefont {C.}~\bibnamefont {Wetterich}},\ }\bibfield  {title} {\bibinfo {title} {{Neutrino Mass Problem and Gauge Hierarchy}},\ }\href {https://doi.org/10.1016/0370-2693(80)90825-4} {\bibfield  {journal} {\bibinfo  {journal} {Phys. Lett. B}\ }\textbf {\bibinfo {volume} {94}},\ \bibinfo {pages} {61} (\bibinfo {year} {1980})}\BibitemShut {NoStop}%
\bibitem [{\citenamefont {Lazarides}\ \emph {et~al.}(1981)\citenamefont {Lazarides}, \citenamefont {Shafi},\ and\ \citenamefont {Wetterich}}]{Lazarides:1980nt}%
  \BibitemOpen
  \bibfield  {author} {\bibinfo {author} {\bibfnamefont {G.}~\bibnamefont {Lazarides}}, \bibinfo {author} {\bibfnamefont {Q.}~\bibnamefont {Shafi}},\ and\ \bibinfo {author} {\bibfnamefont {C.}~\bibnamefont {Wetterich}},\ }\bibfield  {title} {\bibinfo {title} {{Proton Lifetime and Fermion Masses in an SO(10) Model}},\ }\href {https://doi.org/10.1016/0550-3213(81)90354-0} {\bibfield  {journal} {\bibinfo  {journal} {Nucl. Phys. B}\ }\textbf {\bibinfo {volume} {181}},\ \bibinfo {pages} {287} (\bibinfo {year} {1981})}\BibitemShut {NoStop}%
\bibitem [{\citenamefont {Schechter}\ and\ \citenamefont {Valle}(1980)}]{Schechter:1980gr}%
  \BibitemOpen
  \bibfield  {author} {\bibinfo {author} {\bibfnamefont {J.}~\bibnamefont {Schechter}}\ and\ \bibinfo {author} {\bibfnamefont {J.~W.~F.}\ \bibnamefont {Valle}},\ }\bibfield  {title} {\bibinfo {title} {{Neutrino Masses in SU(2) x U(1) Theories}},\ }\href {https://doi.org/10.1103/PhysRevD.22.2227} {\bibfield  {journal} {\bibinfo  {journal} {Phys. Rev. D}\ }\textbf {\bibinfo {volume} {22}},\ \bibinfo {pages} {2227} (\bibinfo {year} {1980})}\BibitemShut {NoStop}%
\bibitem [{\citenamefont {Cheng}\ and\ \citenamefont {Li}(1980)}]{Cheng:1980qt}%
  \BibitemOpen
  \bibfield  {author} {\bibinfo {author} {\bibfnamefont {T.~P.}\ \bibnamefont {Cheng}}\ and\ \bibinfo {author} {\bibfnamefont {L.-F.}\ \bibnamefont {Li}},\ }\bibfield  {title} {\bibinfo {title} {{Neutrino Masses, Mixings and Oscillations in SU(2) x U(1) Models of Electroweak Interactions}},\ }\href {https://doi.org/10.1103/PhysRevD.22.2860} {\bibfield  {journal} {\bibinfo  {journal} {Phys. Rev. D}\ }\textbf {\bibinfo {volume} {22}},\ \bibinfo {pages} {2860} (\bibinfo {year} {1980})}\BibitemShut {NoStop}%
\bibitem [{\citenamefont {Bilenky}\ \emph {et~al.}(1980)\citenamefont {Bilenky}, \citenamefont {Hosek},\ and\ \citenamefont {Petcov}}]{Bilenky:1980cx}%
  \BibitemOpen
  \bibfield  {author} {\bibinfo {author} {\bibfnamefont {S.~M.}\ \bibnamefont {Bilenky}}, \bibinfo {author} {\bibfnamefont {J.}~\bibnamefont {Hosek}},\ and\ \bibinfo {author} {\bibfnamefont {S.~T.}\ \bibnamefont {Petcov}},\ }\bibfield  {title} {\bibinfo {title} {{On Oscillations of Neutrinos with Dirac and Majorana Masses}},\ }\href {https://doi.org/10.1016/0370-2693(80)90927-2} {\bibfield  {journal} {\bibinfo  {journal} {Phys. Lett. B}\ }\textbf {\bibinfo {volume} {94}},\ \bibinfo {pages} {495} (\bibinfo {year} {1980})}\BibitemShut {NoStop}%
\bibitem [{\citenamefont {Wyler}\ and\ \citenamefont {Wolfenstein}(1983)}]{Wyler:1982dd}%
  \BibitemOpen
  \bibfield  {author} {\bibinfo {author} {\bibfnamefont {D.}~\bibnamefont {Wyler}}\ and\ \bibinfo {author} {\bibfnamefont {L.}~\bibnamefont {Wolfenstein}},\ }\bibfield  {title} {\bibinfo {title} {{Massless Neutrinos in Left-Right Symmetric Models}},\ }\href {https://doi.org/10.1016/0550-3213(83)90482-0} {\bibfield  {journal} {\bibinfo  {journal} {Nucl. Phys. B}\ }\textbf {\bibinfo {volume} {218}},\ \bibinfo {pages} {205} (\bibinfo {year} {1983})}\BibitemShut {NoStop}%
\bibitem [{\citenamefont {Mohapatra}\ \emph {et~al.}(2007)\citenamefont {Mohapatra} \emph {et~al.}}]{Mohapatra:2005wg}%
  \BibitemOpen
  \bibfield  {author} {\bibinfo {author} {\bibfnamefont {R.~N.}\ \bibnamefont {Mohapatra}} \emph {et~al.},\ }\bibfield  {title} {\bibinfo {title} {{Theory of Neutrinos: A White Paper}},\ }\href {https://doi.org/10.1088/0034-4885/70/11/R02} {\bibfield  {journal} {\bibinfo  {journal} {Rept. Prog. Phys.}\ }\textbf {\bibinfo {volume} {70}},\ \bibinfo {pages} {1757} (\bibinfo {year} {2007})},\ \Eprint {https://arxiv.org/abs/hep-ph/0510213} {arXiv:hep-ph/0510213} \BibitemShut {NoStop}%
\bibitem [{\citenamefont {Shaposhnikov}(2007)}]{Shaposhnikov:2006nn}%
  \BibitemOpen
  \bibfield  {author} {\bibinfo {author} {\bibfnamefont {M.}~\bibnamefont {Shaposhnikov}},\ }\bibfield  {title} {\bibinfo {title} {{A Possible symmetry of the nuMSM}},\ }\href {https://doi.org/10.1016/j.nuclphysb.2006.11.003} {\bibfield  {journal} {\bibinfo  {journal} {Nucl. Phys. B}\ }\textbf {\bibinfo {volume} {763}},\ \bibinfo {pages} {49} (\bibinfo {year} {2007})},\ \Eprint {https://arxiv.org/abs/hep-ph/0605047} {arXiv:hep-ph/0605047} \BibitemShut {NoStop}%
\bibitem [{\citenamefont {Kersten}\ and\ \citenamefont {Smirnov}(2007)}]{Kersten:2007vk}%
  \BibitemOpen
  \bibfield  {author} {\bibinfo {author} {\bibfnamefont {J.}~\bibnamefont {Kersten}}\ and\ \bibinfo {author} {\bibfnamefont {A.~Y.}\ \bibnamefont {Smirnov}},\ }\bibfield  {title} {\bibinfo {title} {{Right-Handed Neutrinos at CERN LHC and the Mechanism of Neutrino Mass Generation}},\ }\href {https://doi.org/10.1103/PhysRevD.76.073005} {\bibfield  {journal} {\bibinfo  {journal} {Phys. Rev. D}\ }\textbf {\bibinfo {volume} {76}},\ \bibinfo {pages} {073005} (\bibinfo {year} {2007})},\ \Eprint {https://arxiv.org/abs/0705.3221} {arXiv:0705.3221 [hep-ph]} \BibitemShut {NoStop}%
\bibitem [{\citenamefont {Gavela}\ \emph {et~al.}(2009)\citenamefont {Gavela}, \citenamefont {Hambye}, \citenamefont {Hernandez},\ and\ \citenamefont {Hernandez}}]{Gavela:2009cd}%
  \BibitemOpen
  \bibfield  {author} {\bibinfo {author} {\bibfnamefont {M.~B.}\ \bibnamefont {Gavela}}, \bibinfo {author} {\bibfnamefont {T.}~\bibnamefont {Hambye}}, \bibinfo {author} {\bibfnamefont {D.}~\bibnamefont {Hernandez}},\ and\ \bibinfo {author} {\bibfnamefont {P.}~\bibnamefont {Hernandez}},\ }\bibfield  {title} {\bibinfo {title} {{Minimal Flavour Seesaw Models}},\ }\href {https://doi.org/10.1088/1126-6708/2009/09/038} {\bibfield  {journal} {\bibinfo  {journal} {JHEP}\ }\textbf {\bibinfo {volume} {09}},\ \bibinfo {pages} {038}},\ \Eprint {https://arxiv.org/abs/0906.1461} {arXiv:0906.1461 [hep-ph]} \BibitemShut {NoStop}%
\bibitem [{\citenamefont {Ibarra}\ \emph {et~al.}(2010)\citenamefont {Ibarra}, \citenamefont {Molinaro},\ and\ \citenamefont {Petcov}}]{Ibarra:2010xw}%
  \BibitemOpen
  \bibfield  {author} {\bibinfo {author} {\bibfnamefont {A.}~\bibnamefont {Ibarra}}, \bibinfo {author} {\bibfnamefont {E.}~\bibnamefont {Molinaro}},\ and\ \bibinfo {author} {\bibfnamefont {S.~T.}\ \bibnamefont {Petcov}},\ }\bibfield  {title} {\bibinfo {title} {{TeV Scale See-Saw Mechanisms of Neutrino Mass Generation, the Majorana Nature of the Heavy Singlet Neutrinos and $(\beta\beta)_{0\nu}$-Decay}},\ }\href {https://doi.org/10.1007/JHEP09(2010)108} {\bibfield  {journal} {\bibinfo  {journal} {JHEP}\ }\textbf {\bibinfo {volume} {09}},\ \bibinfo {pages} {108}},\ \Eprint {https://arxiv.org/abs/1007.2378} {arXiv:1007.2378 [hep-ph]} \BibitemShut {NoStop}%
\bibitem [{\citenamefont {Antusch}\ and\ \citenamefont {Fischer}(2015)}]{Antusch:2015mia}%
  \BibitemOpen
  \bibfield  {author} {\bibinfo {author} {\bibfnamefont {S.}~\bibnamefont {Antusch}}\ and\ \bibinfo {author} {\bibfnamefont {O.}~\bibnamefont {Fischer}},\ }\bibfield  {title} {\bibinfo {title} {{Testing sterile neutrino extensions of the Standard Model at future lepton colliders}},\ }\href {https://doi.org/10.1007/JHEP05(2015)053} {\bibfield  {journal} {\bibinfo  {journal} {JHEP}\ }\textbf {\bibinfo {volume} {05}},\ \bibinfo {pages} {053}},\ \Eprint {https://arxiv.org/abs/1502.05915} {arXiv:1502.05915 [hep-ph]} \BibitemShut {NoStop}%
\bibitem [{\citenamefont {Moffat}\ \emph {et~al.}(2017)\citenamefont {Moffat}, \citenamefont {Pascoli},\ and\ \citenamefont {Weiland}}]{Moffat:2017feq}%
  \BibitemOpen
  \bibfield  {author} {\bibinfo {author} {\bibfnamefont {K.}~\bibnamefont {Moffat}}, \bibinfo {author} {\bibfnamefont {S.}~\bibnamefont {Pascoli}},\ and\ \bibinfo {author} {\bibfnamefont {C.}~\bibnamefont {Weiland}},\ }\href@noop {} {\bibinfo {title} {{Equivalence between massless neutrinos and lepton number conservation in fermionic singlet extensions of the Standard Model}}} (\bibinfo {year} {2017}),\ \Eprint {https://arxiv.org/abs/1712.07611} {arXiv:1712.07611 [hep-ph]} \BibitemShut {NoStop}%
\bibitem [{\citenamefont {Blennow}\ \emph {et~al.}(2010)\citenamefont {Blennow}, \citenamefont {Fernandez-Martinez}, \citenamefont {Lopez-Pavon},\ and\ \citenamefont {Menendez}}]{Blennow:2010th}%
  \BibitemOpen
  \bibfield  {author} {\bibinfo {author} {\bibfnamefont {M.}~\bibnamefont {Blennow}}, \bibinfo {author} {\bibfnamefont {E.}~\bibnamefont {Fernandez-Martinez}}, \bibinfo {author} {\bibfnamefont {J.}~\bibnamefont {Lopez-Pavon}},\ and\ \bibinfo {author} {\bibfnamefont {J.}~\bibnamefont {Menendez}},\ }\bibfield  {title} {\bibinfo {title} {{Neutrinoless double beta decay in seesaw models}},\ }\href {https://doi.org/10.1007/JHEP07(2010)096} {\bibfield  {journal} {\bibinfo  {journal} {JHEP}\ }\textbf {\bibinfo {volume} {07}},\ \bibinfo {pages} {096}},\ \Eprint {https://arxiv.org/abs/1005.3240} {arXiv:1005.3240 [hep-ph]} \BibitemShut {NoStop}%
\bibitem [{\citenamefont {Mitra}\ \emph {et~al.}(2012)\citenamefont {Mitra}, \citenamefont {Senjanovic},\ and\ \citenamefont {Vissani}}]{Mitra:2011qr}%
  \BibitemOpen
  \bibfield  {author} {\bibinfo {author} {\bibfnamefont {M.}~\bibnamefont {Mitra}}, \bibinfo {author} {\bibfnamefont {G.}~\bibnamefont {Senjanovic}},\ and\ \bibinfo {author} {\bibfnamefont {F.}~\bibnamefont {Vissani}},\ }\bibfield  {title} {\bibinfo {title} {{Neutrinoless Double Beta Decay and Heavy Sterile Neutrinos}},\ }\href {https://doi.org/10.1016/j.nuclphysb.2011.10.035} {\bibfield  {journal} {\bibinfo  {journal} {Nucl. Phys. B}\ }\textbf {\bibinfo {volume} {856}},\ \bibinfo {pages} {26} (\bibinfo {year} {2012})},\ \Eprint {https://arxiv.org/abs/1108.0004} {arXiv:1108.0004 [hep-ph]} \BibitemShut {NoStop}%
\bibitem [{\citenamefont {Atre}\ \emph {et~al.}(2009)\citenamefont {Atre}, \citenamefont {Han}, \citenamefont {Pascoli},\ and\ \citenamefont {Zhang}}]{Atre:2009rg}%
  \BibitemOpen
  \bibfield  {author} {\bibinfo {author} {\bibfnamefont {A.}~\bibnamefont {Atre}}, \bibinfo {author} {\bibfnamefont {T.}~\bibnamefont {Han}}, \bibinfo {author} {\bibfnamefont {S.}~\bibnamefont {Pascoli}},\ and\ \bibinfo {author} {\bibfnamefont {B.}~\bibnamefont {Zhang}},\ }\bibfield  {title} {\bibinfo {title} {{The Search for Heavy Majorana Neutrinos}},\ }\href {https://doi.org/10.1088/1126-6708/2009/05/030} {\bibfield  {journal} {\bibinfo  {journal} {JHEP}\ }\textbf {\bibinfo {volume} {05}},\ \bibinfo {pages} {030}},\ \Eprint {https://arxiv.org/abs/0901.3589} {arXiv:0901.3589 [hep-ph]} \BibitemShut {NoStop}%
\bibitem [{\citenamefont {Aad}\ \emph {et~al.}(2023{\natexlab{a}})\citenamefont {Aad} \emph {et~al.}}]{ATLAS:2023tkz}%
  \BibitemOpen
  \bibfield  {author} {\bibinfo {author} {\bibfnamefont {G.}~\bibnamefont {Aad}} \emph {et~al.} (\bibinfo {collaboration} {ATLAS}),\ }\bibfield  {title} {\bibinfo {title} {{Search for Majorana neutrinos in same-sign WW scattering events from pp collisions at $\sqrt{s}=13$~TeV}},\ }\href {https://doi.org/10.1140/epjc/s10052-023-11915-y} {\bibfield  {journal} {\bibinfo  {journal} {Eur. Phys. J. C}\ }\textbf {\bibinfo {volume} {83}},\ \bibinfo {pages} {824} (\bibinfo {year} {2023}{\natexlab{a}})},\ \Eprint {https://arxiv.org/abs/2305.14931} {arXiv:2305.14931 [hep-ex]} \BibitemShut {NoStop}%
\bibitem [{\citenamefont {Aad}\ \emph {et~al.}(2024)\citenamefont {Aad} \emph {et~al.}}]{ATLAS:2024rzi}%
  \BibitemOpen
  \bibfield  {author} {\bibinfo {author} {\bibfnamefont {G.}~\bibnamefont {Aad}} \emph {et~al.} (\bibinfo {collaboration} {ATLAS}),\ }\bibfield  {title} {\bibinfo {title} {{Search for heavy Majorana neutrinos in e{\ensuremath{\pm}}e{\ensuremath{\pm}} and e{\ensuremath{\pm}}{\ensuremath{\mu}}{\ensuremath{\pm}} final states via WW scattering in pp collisions at s=13 TeV with the ATLAS detector}},\ }\href {https://doi.org/10.1016/j.physletb.2024.138865} {\bibfield  {journal} {\bibinfo  {journal} {Phys. Lett. B}\ }\textbf {\bibinfo {volume} {856}},\ \bibinfo {pages} {138865} (\bibinfo {year} {2024})},\ \Eprint {https://arxiv.org/abs/2403.15016} {arXiv:2403.15016 [hep-ex]} \BibitemShut {NoStop}%
\bibitem [{\citenamefont {Aad}\ \emph {et~al.}(2026)\citenamefont {Aad} \emph {et~al.}}]{ATLAS:2025lva}%
  \BibitemOpen
  \bibfield  {author} {\bibinfo {author} {\bibfnamefont {G.}~\bibnamefont {Aad}} \emph {et~al.} (\bibinfo {collaboration} {ATLAS}),\ }\bibfield  {title} {\bibinfo {title} {{Search for heavy neutral leptons in decays of W bosons produced in 13 TeV $pp$ collisions using prompt signatures in the ATLAS detector}},\ }\href {https://doi.org/10.1140/epjc/s10052-025-15191-w} {\bibfield  {journal} {\bibinfo  {journal} {Eur. Phys. J. C}\ }\textbf {\bibinfo {volume} {86}},\ \bibinfo {pages} {153} (\bibinfo {year} {2026})},\ \Eprint {https://arxiv.org/abs/2508.20929} {arXiv:2508.20929 [hep-ex]} \BibitemShut {NoStop}%
\bibitem [{\citenamefont {del Aguila}\ \emph {et~al.}(2008)\citenamefont {del Aguila}, \citenamefont {de~Blas},\ and\ \citenamefont {Perez-Victoria}}]{delAguila:2008pw}%
  \BibitemOpen
  \bibfield  {author} {\bibinfo {author} {\bibfnamefont {F.}~\bibnamefont {del Aguila}}, \bibinfo {author} {\bibfnamefont {J.}~\bibnamefont {de~Blas}},\ and\ \bibinfo {author} {\bibfnamefont {M.}~\bibnamefont {Perez-Victoria}},\ }\bibfield  {title} {\bibinfo {title} {{Effects of new leptons in Electroweak Precision Data}},\ }\href {https://doi.org/10.1103/PhysRevD.78.013010} {\bibfield  {journal} {\bibinfo  {journal} {Phys. Rev. D}\ }\textbf {\bibinfo {volume} {78}},\ \bibinfo {pages} {013010} (\bibinfo {year} {2008})},\ \Eprint {https://arxiv.org/abs/0803.4008} {arXiv:0803.4008 [hep-ph]} \BibitemShut {NoStop}%
\bibitem [{\citenamefont {Akhmedov}\ \emph {et~al.}(2013)\citenamefont {Akhmedov}, \citenamefont {Kartavtsev}, \citenamefont {Lindner}, \citenamefont {Michaels},\ and\ \citenamefont {Smirnov}}]{Akhmedov:2013hec}%
  \BibitemOpen
  \bibfield  {author} {\bibinfo {author} {\bibfnamefont {E.}~\bibnamefont {Akhmedov}}, \bibinfo {author} {\bibfnamefont {A.}~\bibnamefont {Kartavtsev}}, \bibinfo {author} {\bibfnamefont {M.}~\bibnamefont {Lindner}}, \bibinfo {author} {\bibfnamefont {L.}~\bibnamefont {Michaels}},\ and\ \bibinfo {author} {\bibfnamefont {J.}~\bibnamefont {Smirnov}},\ }\bibfield  {title} {\bibinfo {title} {{Improving Electro-Weak Fits with TeV-scale Sterile Neutrinos}},\ }\href {https://doi.org/10.1007/JHEP05(2013)081} {\bibfield  {journal} {\bibinfo  {journal} {JHEP}\ }\textbf {\bibinfo {volume} {05}},\ \bibinfo {pages} {081}},\ \Eprint {https://arxiv.org/abs/1302.1872} {arXiv:1302.1872 [hep-ph]} \BibitemShut {NoStop}%
\bibitem [{\citenamefont {de~Blas}(2013)}]{deBlas:2013gla}%
  \BibitemOpen
  \bibfield  {author} {\bibinfo {author} {\bibfnamefont {J.}~\bibnamefont {de~Blas}},\ }\bibfield  {title} {\bibinfo {title} {{Electroweak limits on physics beyond the Standard Model}},\ }\href {https://doi.org/10.1051/epjconf/20136019008} {\bibfield  {journal} {\bibinfo  {journal} {EPJ Web Conf.}\ }\textbf {\bibinfo {volume} {60}},\ \bibinfo {pages} {19008} (\bibinfo {year} {2013})},\ \Eprint {https://arxiv.org/abs/1307.6173} {arXiv:1307.6173 [hep-ph]} \BibitemShut {NoStop}%
\bibitem [{\citenamefont {Basso}\ \emph {et~al.}(2014)\citenamefont {Basso}, \citenamefont {Fischer},\ and\ \citenamefont {van~der Bij}}]{Basso:2013jka}%
  \BibitemOpen
  \bibfield  {author} {\bibinfo {author} {\bibfnamefont {L.}~\bibnamefont {Basso}}, \bibinfo {author} {\bibfnamefont {O.}~\bibnamefont {Fischer}},\ and\ \bibinfo {author} {\bibfnamefont {J.~J.}\ \bibnamefont {van~der Bij}},\ }\bibfield  {title} {\bibinfo {title} {{Precision tests of unitarity in leptonic mixing}},\ }\href {https://doi.org/10.1209/0295-5075/105/11001} {\bibfield  {journal} {\bibinfo  {journal} {EPL}\ }\textbf {\bibinfo {volume} {105}},\ \bibinfo {pages} {11001} (\bibinfo {year} {2014})},\ \Eprint {https://arxiv.org/abs/1310.2057} {arXiv:1310.2057 [hep-ph]} \BibitemShut {NoStop}%
\bibitem [{\citenamefont {Antusch}\ and\ \citenamefont {Fischer}(2014)}]{Antusch:2014woa}%
  \BibitemOpen
  \bibfield  {author} {\bibinfo {author} {\bibfnamefont {S.}~\bibnamefont {Antusch}}\ and\ \bibinfo {author} {\bibfnamefont {O.}~\bibnamefont {Fischer}},\ }\bibfield  {title} {\bibinfo {title} {{Non-unitarity of the leptonic mixing matrix: Present bounds and future sensitivities}},\ }\href {https://doi.org/10.1007/JHEP10(2014)094} {\bibfield  {journal} {\bibinfo  {journal} {JHEP}\ }\textbf {\bibinfo {volume} {10}},\ \bibinfo {pages} {094}},\ \Eprint {https://arxiv.org/abs/1407.6607} {arXiv:1407.6607 [hep-ph]} \BibitemShut {NoStop}%
\bibitem [{\citenamefont {Chrzaszcz}\ \emph {et~al.}(2020)\citenamefont {Chrzaszcz}, \citenamefont {Drewes}, \citenamefont {Gonzalo}, \citenamefont {Harz}, \citenamefont {Krishnamurthy},\ and\ \citenamefont {Weniger}}]{Chrzaszcz:2019inj}%
  \BibitemOpen
  \bibfield  {author} {\bibinfo {author} {\bibfnamefont {M.}~\bibnamefont {Chrzaszcz}}, \bibinfo {author} {\bibfnamefont {M.}~\bibnamefont {Drewes}}, \bibinfo {author} {\bibfnamefont {T.~E.}\ \bibnamefont {Gonzalo}}, \bibinfo {author} {\bibfnamefont {J.}~\bibnamefont {Harz}}, \bibinfo {author} {\bibfnamefont {S.}~\bibnamefont {Krishnamurthy}},\ and\ \bibinfo {author} {\bibfnamefont {C.}~\bibnamefont {Weniger}},\ }\bibfield  {title} {\bibinfo {title} {{A frequentist analysis of three right-handed neutrinos with GAMBIT}},\ }\href {https://doi.org/10.1140/epjc/s10052-020-8073-9} {\bibfield  {journal} {\bibinfo  {journal} {Eur. Phys. J. C}\ }\textbf {\bibinfo {volume} {80}},\ \bibinfo {pages} {569} (\bibinfo {year} {2020})},\ \Eprint {https://arxiv.org/abs/1908.02302} {arXiv:1908.02302 [hep-ph]} \BibitemShut {NoStop}%
\bibitem [{\citenamefont {Bryman}\ \emph {et~al.}(2022)\citenamefont {Bryman}, \citenamefont {Cirigliano}, \citenamefont {Crivellin},\ and\ \citenamefont {Inguglia}}]{Bryman:2021teu}%
  \BibitemOpen
  \bibfield  {author} {\bibinfo {author} {\bibfnamefont {D.}~\bibnamefont {Bryman}}, \bibinfo {author} {\bibfnamefont {V.}~\bibnamefont {Cirigliano}}, \bibinfo {author} {\bibfnamefont {A.}~\bibnamefont {Crivellin}},\ and\ \bibinfo {author} {\bibfnamefont {G.}~\bibnamefont {Inguglia}},\ }\bibfield  {title} {\bibinfo {title} {{Testing Lepton Flavor Universality with Pion, Kaon, Tau, and Beta Decays}},\ }\href {https://doi.org/10.1146/annurev-nucl-110121-051223} {\bibfield  {journal} {\bibinfo  {journal} {Ann. Rev. Nucl. Part. Sci.}\ }\textbf {\bibinfo {volume} {72}},\ \bibinfo {pages} {69} (\bibinfo {year} {2022})},\ \Eprint {https://arxiv.org/abs/2111.05338} {arXiv:2111.05338 [hep-ph]} \BibitemShut {NoStop}%
\bibitem [{\citenamefont {Blennow}\ \emph {et~al.}(2023)\citenamefont {Blennow}, \citenamefont {Fern{\'a}ndez-Mart{\'\i}nez}, \citenamefont {Hern{\'a}ndez-Garc{\'\i}a}, \citenamefont {L{\'o}pez-Pav{\'o}n}, \citenamefont {Marcano},\ and\ \citenamefont {Naredo-Tuero}}]{Blennow:2023mqx}%
  \BibitemOpen
  \bibfield  {author} {\bibinfo {author} {\bibfnamefont {M.}~\bibnamefont {Blennow}}, \bibinfo {author} {\bibfnamefont {E.}~\bibnamefont {Fern{\'a}ndez-Mart{\'\i}nez}}, \bibinfo {author} {\bibfnamefont {J.}~\bibnamefont {Hern{\'a}ndez-Garc{\'\i}a}}, \bibinfo {author} {\bibfnamefont {J.}~\bibnamefont {L{\'o}pez-Pav{\'o}n}}, \bibinfo {author} {\bibfnamefont {X.}~\bibnamefont {Marcano}},\ and\ \bibinfo {author} {\bibfnamefont {D.}~\bibnamefont {Naredo-Tuero}},\ }\bibfield  {title} {\bibinfo {title} {{Bounds on lepton non-unitarity and heavy neutrino mixing}},\ }\href {https://doi.org/10.1007/JHEP08(2023)030} {\bibfield  {journal} {\bibinfo  {journal} {JHEP}\ }\textbf {\bibinfo {volume} {08}},\ \bibinfo {pages} {030}},\ \Eprint {https://arxiv.org/abs/2306.01040} {arXiv:2306.01040 [hep-ph]} \BibitemShut {NoStop}%
\bibitem [{\citenamefont {Keung}\ and\ \citenamefont {Senjanovic}(1983)}]{Keung:1983uu}%
  \BibitemOpen
  \bibfield  {author} {\bibinfo {author} {\bibfnamefont {W.-Y.}\ \bibnamefont {Keung}}\ and\ \bibinfo {author} {\bibfnamefont {G.}~\bibnamefont {Senjanovic}},\ }\bibfield  {title} {\bibinfo {title} {{Majorana Neutrinos and the Production of the Right-handed Charged Gauge Boson}},\ }\href {https://doi.org/10.1103/PhysRevLett.50.1427} {\bibfield  {journal} {\bibinfo  {journal} {Phys. Rev. Lett.}\ }\textbf {\bibinfo {volume} {50}},\ \bibinfo {pages} {1427} (\bibinfo {year} {1983})}\BibitemShut {NoStop}%
\bibitem [{\citenamefont {Ilakovac}\ and\ \citenamefont {Pilaftsis}(1995)}]{Ilakovac:1994kj}%
  \BibitemOpen
  \bibfield  {author} {\bibinfo {author} {\bibfnamefont {A.}~\bibnamefont {Ilakovac}}\ and\ \bibinfo {author} {\bibfnamefont {A.}~\bibnamefont {Pilaftsis}},\ }\bibfield  {title} {\bibinfo {title} {{Flavor violating charged lepton decays in seesaw-type models}},\ }\href {https://doi.org/10.1016/0550-3213(94)00567-X} {\bibfield  {journal} {\bibinfo  {journal} {Nucl. Phys. B}\ }\textbf {\bibinfo {volume} {437}},\ \bibinfo {pages} {491} (\bibinfo {year} {1995})},\ \Eprint {https://arxiv.org/abs/hep-ph/9403398} {arXiv:hep-ph/9403398} \BibitemShut {NoStop}%
\bibitem [{\citenamefont {Alonso}\ \emph {et~al.}(2013)\citenamefont {Alonso}, \citenamefont {Dhen}, \citenamefont {Gavela},\ and\ \citenamefont {Hambye}}]{Alonso:2012ji}%
  \BibitemOpen
  \bibfield  {author} {\bibinfo {author} {\bibfnamefont {R.}~\bibnamefont {Alonso}}, \bibinfo {author} {\bibfnamefont {M.}~\bibnamefont {Dhen}}, \bibinfo {author} {\bibfnamefont {M.~B.}\ \bibnamefont {Gavela}},\ and\ \bibinfo {author} {\bibfnamefont {T.}~\bibnamefont {Hambye}},\ }\bibfield  {title} {\bibinfo {title} {{Muon conversion to electron in nuclei in type-I seesaw models}},\ }\href {https://doi.org/10.1007/JHEP01(2013)118} {\bibfield  {journal} {\bibinfo  {journal} {JHEP}\ }\textbf {\bibinfo {volume} {01}},\ \bibinfo {pages} {118}},\ \Eprint {https://arxiv.org/abs/1209.2679} {arXiv:1209.2679 [hep-ph]} \BibitemShut {NoStop}%
\bibitem [{\citenamefont {Daum}\ \emph {et~al.}(1987)\citenamefont {Daum}, \citenamefont {Jost}, \citenamefont {Marshall}, \citenamefont {Minehart}, \citenamefont {Stephens},\ and\ \citenamefont {Ziock}}]{Daum:1987bg}%
  \BibitemOpen
  \bibfield  {author} {\bibinfo {author} {\bibfnamefont {M.}~\bibnamefont {Daum}}, \bibinfo {author} {\bibfnamefont {B.}~\bibnamefont {Jost}}, \bibinfo {author} {\bibfnamefont {R.~M.}\ \bibnamefont {Marshall}}, \bibinfo {author} {\bibfnamefont {R.~C.}\ \bibnamefont {Minehart}}, \bibinfo {author} {\bibfnamefont {W.~A.}\ \bibnamefont {Stephens}},\ and\ \bibinfo {author} {\bibfnamefont {K.~O.~H.}\ \bibnamefont {Ziock}},\ }\bibfield  {title} {\bibinfo {title} {{Search for Admixtures of Massive Neutrinos in the Decay $\pi^+ \to \mu^+$ Neutrino}},\ }\href {https://doi.org/10.1103/PhysRevD.36.2624} {\bibfield  {journal} {\bibinfo  {journal} {Phys. Rev. D}\ }\textbf {\bibinfo {volume} {36}},\ \bibinfo {pages} {2624} (\bibinfo {year} {1987})}\BibitemShut {NoStop}%
\bibitem [{\citenamefont {Aguilar-Arevalo}\ \emph {et~al.}(2019)\citenamefont {Aguilar-Arevalo} \emph {et~al.}}]{PIENU:2019usb}%
  \BibitemOpen
  \bibfield  {author} {\bibinfo {author} {\bibfnamefont {A.}~\bibnamefont {Aguilar-Arevalo}} \emph {et~al.} (\bibinfo {collaboration} {PIENU}),\ }\bibfield  {title} {\bibinfo {title} {{Search for heavy neutrinos in $\pi \to \mu\nu$ decay}},\ }\href {https://doi.org/10.1016/j.physletb.2019.134980} {\bibfield  {journal} {\bibinfo  {journal} {Phys. Lett. B}\ }\textbf {\bibinfo {volume} {798}},\ \bibinfo {pages} {134980} (\bibinfo {year} {2019})},\ \Eprint {https://arxiv.org/abs/1904.03269} {arXiv:1904.03269 [hep-ex]} \BibitemShut {NoStop}%
\bibitem [{\citenamefont {Bernardi}\ \emph {et~al.}(1986)\citenamefont {Bernardi} \emph {et~al.}}]{Bernardi:1985ny}%
  \BibitemOpen
  \bibfield  {author} {\bibinfo {author} {\bibfnamefont {G.}~\bibnamefont {Bernardi}} \emph {et~al.},\ }\bibfield  {title} {\bibinfo {title} {{Search for Neutrino Decay}},\ }\href {https://doi.org/10.1016/0370-2693(86)91602-3} {\bibfield  {journal} {\bibinfo  {journal} {Phys. Lett. B}\ }\textbf {\bibinfo {volume} {166}},\ \bibinfo {pages} {479} (\bibinfo {year} {1986})}\BibitemShut {NoStop}%
\bibitem [{\citenamefont {Bernardi}\ \emph {et~al.}(1988)\citenamefont {Bernardi} \emph {et~al.}}]{Bernardi:1987ek}%
  \BibitemOpen
  \bibfield  {author} {\bibinfo {author} {\bibfnamefont {G.}~\bibnamefont {Bernardi}} \emph {et~al.},\ }\bibfield  {title} {\bibinfo {title} {{FURTHER LIMITS ON HEAVY NEUTRINO COUPLINGS}},\ }\href {https://doi.org/10.1016/0370-2693(88)90563-1} {\bibfield  {journal} {\bibinfo  {journal} {Phys. Lett. B}\ }\textbf {\bibinfo {volume} {203}},\ \bibinfo {pages} {332} (\bibinfo {year} {1988})}\BibitemShut {NoStop}%
\bibitem [{\citenamefont {Abe}\ \emph {et~al.}(2019)\citenamefont {Abe} \emph {et~al.}}]{T2K:2019jwa}%
  \BibitemOpen
  \bibfield  {author} {\bibinfo {author} {\bibfnamefont {K.}~\bibnamefont {Abe}} \emph {et~al.} (\bibinfo {collaboration} {T2K}),\ }\bibfield  {title} {\bibinfo {title} {{Search for heavy neutrinos with the T2K near detector ND280}},\ }\href {https://doi.org/10.1103/PhysRevD.100.052006} {\bibfield  {journal} {\bibinfo  {journal} {Phys. Rev. D}\ }\textbf {\bibinfo {volume} {100}},\ \bibinfo {pages} {052006} (\bibinfo {year} {2019})},\ \Eprint {https://arxiv.org/abs/1902.07598} {arXiv:1902.07598 [hep-ex]} \BibitemShut {NoStop}%
\bibitem [{\citenamefont {Kelly}\ and\ \citenamefont {Machado}(2021)}]{Kelly:2021xbv}%
  \BibitemOpen
  \bibfield  {author} {\bibinfo {author} {\bibfnamefont {K.~J.}\ \bibnamefont {Kelly}}\ and\ \bibinfo {author} {\bibfnamefont {P.~A.~N.}\ \bibnamefont {Machado}},\ }\bibfield  {title} {\bibinfo {title} {{MicroBooNE experiment, NuMI absorber, and heavy neutral leptons}},\ }\href {https://doi.org/10.1103/PhysRevD.104.055015} {\bibfield  {journal} {\bibinfo  {journal} {Phys. Rev. D}\ }\textbf {\bibinfo {volume} {104}},\ \bibinfo {pages} {055015} (\bibinfo {year} {2021})},\ \Eprint {https://arxiv.org/abs/2106.06548} {arXiv:2106.06548 [hep-ph]} \BibitemShut {NoStop}%
\bibitem [{\citenamefont {Arg{\"u}elles}\ \emph {et~al.}(2022)\citenamefont {Arg{\"u}elles}, \citenamefont {Foppiani},\ and\ \citenamefont {Hostert}}]{Arguelles:2021dqn}%
  \BibitemOpen
  \bibfield  {author} {\bibinfo {author} {\bibfnamefont {C.~A.}\ \bibnamefont {Arg{\"u}elles}}, \bibinfo {author} {\bibfnamefont {N.}~\bibnamefont {Foppiani}},\ and\ \bibinfo {author} {\bibfnamefont {M.}~\bibnamefont {Hostert}},\ }\bibfield  {title} {\bibinfo {title} {{Heavy neutral leptons below the kaon mass at hodoscopic neutrino detectors}},\ }\href {https://doi.org/10.1103/PhysRevD.105.095006} {\bibfield  {journal} {\bibinfo  {journal} {Phys. Rev. D}\ }\textbf {\bibinfo {volume} {105}},\ \bibinfo {pages} {095006} (\bibinfo {year} {2022})},\ \Eprint {https://arxiv.org/abs/2109.03831} {arXiv:2109.03831 [hep-ph]} \BibitemShut {NoStop}%
\bibitem [{\citenamefont {Hayano}\ \emph {et~al.}(1982)\citenamefont {Hayano} \emph {et~al.}}]{Hayano:1982wu}%
  \BibitemOpen
  \bibfield  {author} {\bibinfo {author} {\bibfnamefont {R.~S.}\ \bibnamefont {Hayano}} \emph {et~al.},\ }\bibfield  {title} {\bibinfo {title} {{HEAVY NEUTRINO SEARCH USING K(mu2) DECAY}},\ }\href {https://doi.org/10.1103/PhysRevLett.49.1305} {\bibfield  {journal} {\bibinfo  {journal} {Phys. Rev. Lett.}\ }\textbf {\bibinfo {volume} {49}},\ \bibinfo {pages} {1305} (\bibinfo {year} {1982})}\BibitemShut {NoStop}%
\bibitem [{\citenamefont {Yamazaki}\ \emph {et~al.}(1984)\citenamefont {Yamazaki} \emph {et~al.}}]{Yamazaki:1984sj}%
  \BibitemOpen
  \bibfield  {author} {\bibinfo {author} {\bibfnamefont {T.}~\bibnamefont {Yamazaki}} \emph {et~al.},\ }\bibfield  {title} {\bibinfo {title} {{Search for Heavy Neutrinos in Kaon Decay}},\ }\href@noop {} {\bibfield  {journal} {\bibinfo  {journal} {Conf. Proc. C}\ }\textbf {\bibinfo {volume} {840719}},\ \bibinfo {pages} {262} (\bibinfo {year} {1984})}\BibitemShut {NoStop}%
\bibitem [{\citenamefont {Artamonov}\ \emph {et~al.}(2009)\citenamefont {Artamonov} \emph {et~al.}}]{BNL-E949:2009dza}%
  \BibitemOpen
  \bibfield  {author} {\bibinfo {author} {\bibfnamefont {A.~V.}\ \bibnamefont {Artamonov}} \emph {et~al.} (\bibinfo {collaboration} {BNL-E949}),\ }\bibfield  {title} {\bibinfo {title} {{Study of the decay $K^+\to\pi^+\nu \bar\nu$ in the momentum region $140 < P_\pi < 199$ MeV/c}},\ }\href {https://doi.org/10.1103/PhysRevD.79.092004} {\bibfield  {journal} {\bibinfo  {journal} {Phys. Rev. D}\ }\textbf {\bibinfo {volume} {79}},\ \bibinfo {pages} {092004} (\bibinfo {year} {2009})},\ \Eprint {https://arxiv.org/abs/0903.0030} {arXiv:0903.0030 [hep-ex]} \BibitemShut {NoStop}%
\bibitem [{\citenamefont {Cortina~Gil}\ \emph {et~al.}(2021)\citenamefont {Cortina~Gil} \emph {et~al.}}]{NA62:2021bji}%
  \BibitemOpen
  \bibfield  {author} {\bibinfo {author} {\bibfnamefont {E.}~\bibnamefont {Cortina~Gil}} \emph {et~al.} (\bibinfo {collaboration} {NA62}),\ }\bibfield  {title} {\bibinfo {title} {{Search for $K^+$ decays to a muon and invisible particles}},\ }\href {https://doi.org/10.1016/j.physletb.2021.136259} {\bibfield  {journal} {\bibinfo  {journal} {Phys. Lett. B}\ }\textbf {\bibinfo {volume} {816}},\ \bibinfo {pages} {136259} (\bibinfo {year} {2021})},\ \Eprint {https://arxiv.org/abs/2101.12304} {arXiv:2101.12304 [hep-ex]} \BibitemShut {NoStop}%
\bibitem [{\citenamefont {Cooper-Sarkar}\ \emph {et~al.}(1985)\citenamefont {Cooper-Sarkar} \emph {et~al.}}]{WA66:1985mfx}%
  \BibitemOpen
  \bibfield  {author} {\bibinfo {author} {\bibfnamefont {A.~M.}\ \bibnamefont {Cooper-Sarkar}} \emph {et~al.} (\bibinfo {collaboration} {WA66}),\ }\bibfield  {title} {\bibinfo {title} {{Search for Heavy Neutrino Decays in the {BEBC} Beam Dump Experiment}},\ }\href {https://doi.org/10.1016/0370-2693(85)91493-5} {\bibfield  {journal} {\bibinfo  {journal} {Phys. Lett. B}\ }\textbf {\bibinfo {volume} {160}},\ \bibinfo {pages} {207} (\bibinfo {year} {1985})}\BibitemShut {NoStop}%
\bibitem [{\citenamefont {Vaitaitis}\ \emph {et~al.}(1999)\citenamefont {Vaitaitis} \emph {et~al.}}]{NuTeV:1999kej}%
  \BibitemOpen
  \bibfield  {author} {\bibinfo {author} {\bibfnamefont {A.}~\bibnamefont {Vaitaitis}} \emph {et~al.} (\bibinfo {collaboration} {NuTeV, E815}),\ }\bibfield  {title} {\bibinfo {title} {{Search for neutral heavy leptons in a high-energy neutrino beam}},\ }\href {https://doi.org/10.1103/PhysRevLett.83.4943} {\bibfield  {journal} {\bibinfo  {journal} {Phys. Rev. Lett.}\ }\textbf {\bibinfo {volume} {83}},\ \bibinfo {pages} {4943} (\bibinfo {year} {1999})},\ \Eprint {https://arxiv.org/abs/hep-ex/9908011} {arXiv:hep-ex/9908011} \BibitemShut {NoStop}%
\bibitem [{\citenamefont {Bergsma}\ \emph {et~al.}(1986)\citenamefont {Bergsma} \emph {et~al.}}]{CHARM:1985nku}%
  \BibitemOpen
  \bibfield  {author} {\bibinfo {author} {\bibfnamefont {F.}~\bibnamefont {Bergsma}} \emph {et~al.} (\bibinfo {collaboration} {CHARM}),\ }\bibfield  {title} {\bibinfo {title} {{A Search for Decays of Heavy Neutrinos in the Mass Range 0.5-{GeV} to 2.8-{GeV}}},\ }\href {https://doi.org/10.1016/0370-2693(86)91601-1} {\bibfield  {journal} {\bibinfo  {journal} {Phys. Lett. B}\ }\textbf {\bibinfo {volume} {166}},\ \bibinfo {pages} {473} (\bibinfo {year} {1986})}\BibitemShut {NoStop}%
\bibitem [{\citenamefont {Aad}\ \emph {et~al.}(2019)\citenamefont {Aad} \emph {et~al.}}]{ATLAS:2019kpx}%
  \BibitemOpen
  \bibfield  {author} {\bibinfo {author} {\bibfnamefont {G.}~\bibnamefont {Aad}} \emph {et~al.} (\bibinfo {collaboration} {ATLAS}),\ }\bibfield  {title} {\bibinfo {title} {{Search for heavy neutral leptons in decays of $W$ bosons produced in 13 TeV $pp$ collisions using prompt and displaced signatures with the ATLAS detector}},\ }\href {https://doi.org/10.1007/JHEP10(2019)265} {\bibfield  {journal} {\bibinfo  {journal} {JHEP}\ }\textbf {\bibinfo {volume} {10}},\ \bibinfo {pages} {265}},\ \Eprint {https://arxiv.org/abs/1905.09787} {arXiv:1905.09787 [hep-ex]} \BibitemShut {NoStop}%
\bibitem [{\citenamefont {Aad}\ \emph {et~al.}(2023{\natexlab{b}})\citenamefont {Aad} \emph {et~al.}}]{ATLAS:2022atq}%
  \BibitemOpen
  \bibfield  {author} {\bibinfo {author} {\bibfnamefont {G.}~\bibnamefont {Aad}} \emph {et~al.} (\bibinfo {collaboration} {ATLAS}),\ }\bibfield  {title} {\bibinfo {title} {{Search for Heavy Neutral Leptons in Decays of W Bosons Using a Dilepton Displaced Vertex in s=13{\,}{\,}TeV pp Collisions with the ATLAS Detector}},\ }\href {https://doi.org/10.1103/PhysRevLett.131.061803} {\bibfield  {journal} {\bibinfo  {journal} {Phys. Rev. Lett.}\ }\textbf {\bibinfo {volume} {131}},\ \bibinfo {pages} {061803} (\bibinfo {year} {2023}{\natexlab{b}})},\ \Eprint {https://arxiv.org/abs/2204.11988} {arXiv:2204.11988 [hep-ex]} \BibitemShut {NoStop}%
\bibitem [{\citenamefont {Tumasyan}\ \emph {et~al.}(2022)\citenamefont {Tumasyan} \emph {et~al.}}]{CMS:2022fut}%
  \BibitemOpen
  \bibfield  {author} {\bibinfo {author} {\bibfnamefont {A.}~\bibnamefont {Tumasyan}} \emph {et~al.} (\bibinfo {collaboration} {CMS}),\ }\bibfield  {title} {\bibinfo {title} {{Search for long-lived heavy neutral leptons with displaced vertices in proton-proton collisions at $ \sqrt{\mathrm{s}} $ =13 TeV}},\ }\href {https://doi.org/10.1007/JHEP07(2022)081} {\bibfield  {journal} {\bibinfo  {journal} {JHEP}\ }\textbf {\bibinfo {volume} {07}},\ \bibinfo {pages} {081}},\ \Eprint {https://arxiv.org/abs/2201.05578} {arXiv:2201.05578 [hep-ex]} \BibitemShut {NoStop}%
\bibitem [{\citenamefont {Abreu}\ \emph {et~al.}(1997)\citenamefont {Abreu} \emph {et~al.}}]{DELPHI:1996qcc}%
  \BibitemOpen
  \bibfield  {author} {\bibinfo {author} {\bibfnamefont {P.}~\bibnamefont {Abreu}} \emph {et~al.} (\bibinfo {collaboration} {DELPHI}),\ }\bibfield  {title} {\bibinfo {title} {{Search for neutral heavy leptons produced in Z decays}},\ }\href {https://doi.org/10.1007/s002880050370} {\bibfield  {journal} {\bibinfo  {journal} {Z. Phys. C}\ }\textbf {\bibinfo {volume} {74}},\ \bibinfo {pages} {57} (\bibinfo {year} {1997})},\ \bibinfo {note} {[Erratum: Z.Phys.C 75, 580 (1997)]}\BibitemShut {NoStop}%
\bibitem [{\citenamefont {Fernandez-Martinez}\ \emph {et~al.}(2016)\citenamefont {Fernandez-Martinez}, \citenamefont {Hernandez-Garcia},\ and\ \citenamefont {Lopez-Pavon}}]{Fernandez-Martinez:2016lgt}%
  \BibitemOpen
  \bibfield  {author} {\bibinfo {author} {\bibfnamefont {E.}~\bibnamefont {Fernandez-Martinez}}, \bibinfo {author} {\bibfnamefont {J.}~\bibnamefont {Hernandez-Garcia}},\ and\ \bibinfo {author} {\bibfnamefont {J.}~\bibnamefont {Lopez-Pavon}},\ }\bibfield  {title} {\bibinfo {title} {{Global constraints on heavy neutrino mixing}},\ }\href {https://doi.org/10.1007/JHEP08(2016)033} {\bibfield  {journal} {\bibinfo  {journal} {JHEP}\ }\textbf {\bibinfo {volume} {08}},\ \bibinfo {pages} {033}},\ \Eprint {https://arxiv.org/abs/1605.08774} {arXiv:1605.08774 [hep-ph]} \BibitemShut {NoStop}%
\bibitem [{\citenamefont {Abdullahi}\ \emph {et~al.}(2023)\citenamefont {Abdullahi} \emph {et~al.}}]{Abdullahi:2022jlv}%
  \BibitemOpen
  \bibfield  {author} {\bibinfo {author} {\bibfnamefont {A.~M.}\ \bibnamefont {Abdullahi}} \emph {et~al.},\ }\bibfield  {title} {\bibinfo {title} {{The present and future status of heavy neutral leptons}},\ }\href {https://doi.org/10.1088/1361-6471/ac98f9} {\bibfield  {journal} {\bibinfo  {journal} {J. Phys. G}\ }\textbf {\bibinfo {volume} {50}},\ \bibinfo {pages} {020501} (\bibinfo {year} {2023})},\ \Eprint {https://arxiv.org/abs/2203.08039} {arXiv:2203.08039 [hep-ph]} \BibitemShut {NoStop}%
\bibitem [{\citenamefont {Krasnov}(2019)}]{Krasnov:2019kdc}%
  \BibitemOpen
  \bibfield  {author} {\bibinfo {author} {\bibfnamefont {I.}~\bibnamefont {Krasnov}},\ }\bibfield  {title} {\bibinfo {title} {{DUNE prospects in the search for sterile neutrinos}},\ }\href {https://doi.org/10.1103/PhysRevD.100.075023} {\bibfield  {journal} {\bibinfo  {journal} {Phys. Rev. D}\ }\textbf {\bibinfo {volume} {100}},\ \bibinfo {pages} {075023} (\bibinfo {year} {2019})},\ \Eprint {https://arxiv.org/abs/1902.06099} {arXiv:1902.06099 [hep-ph]} \BibitemShut {NoStop}%
\bibitem [{\citenamefont {Ballett}\ \emph {et~al.}(2020)\citenamefont {Ballett}, \citenamefont {Boschi},\ and\ \citenamefont {Pascoli}}]{Ballett:2019bgd}%
  \BibitemOpen
  \bibfield  {author} {\bibinfo {author} {\bibfnamefont {P.}~\bibnamefont {Ballett}}, \bibinfo {author} {\bibfnamefont {T.}~\bibnamefont {Boschi}},\ and\ \bibinfo {author} {\bibfnamefont {S.}~\bibnamefont {Pascoli}},\ }\bibfield  {title} {\bibinfo {title} {{Heavy Neutral Leptons from low-scale seesaws at the DUNE Near Detector}},\ }\href {https://doi.org/10.1007/JHEP03(2020)111} {\bibfield  {journal} {\bibinfo  {journal} {JHEP}\ }\textbf {\bibinfo {volume} {03}},\ \bibinfo {pages} {111}},\ \Eprint {https://arxiv.org/abs/1905.00284} {arXiv:1905.00284 [hep-ph]} \BibitemShut {NoStop}%
\bibitem [{\citenamefont {Carbajal}\ and\ \citenamefont {Gago}(2024)}]{Carbajal:2022zlp}%
  \BibitemOpen
  \bibfield  {author} {\bibinfo {author} {\bibfnamefont {S.}~\bibnamefont {Carbajal}}\ and\ \bibinfo {author} {\bibfnamefont {A.~M.}\ \bibnamefont {Gago}},\ }\bibfield  {title} {\bibinfo {title} {{Indirect search of heavy neutral leptons using the DUNE near detector}},\ }\href {https://doi.org/10.3389/fphy.2024.1398070} {\bibfield  {journal} {\bibinfo  {journal} {Front. in Phys.}\ }\textbf {\bibinfo {volume} {12}},\ \bibinfo {pages} {1398070} (\bibinfo {year} {2024})},\ \Eprint {https://arxiv.org/abs/2202.09217} {arXiv:2202.09217 [hep-ph]} \BibitemShut {NoStop}%
\bibitem [{\citenamefont {Feng}\ \emph {et~al.}(2018)\citenamefont {Feng}, \citenamefont {Galon}, \citenamefont {Kling},\ and\ \citenamefont {Trojanowski}}]{Feng:2017uoz}%
  \BibitemOpen
  \bibfield  {author} {\bibinfo {author} {\bibfnamefont {J.~L.}\ \bibnamefont {Feng}}, \bibinfo {author} {\bibfnamefont {I.}~\bibnamefont {Galon}}, \bibinfo {author} {\bibfnamefont {F.}~\bibnamefont {Kling}},\ and\ \bibinfo {author} {\bibfnamefont {S.}~\bibnamefont {Trojanowski}},\ }\bibfield  {title} {\bibinfo {title} {{ForwArd Search ExpeRiment at the LHC}},\ }\href {https://doi.org/10.1103/PhysRevD.97.035001} {\bibfield  {journal} {\bibinfo  {journal} {Phys. Rev. D}\ }\textbf {\bibinfo {volume} {97}},\ \bibinfo {pages} {035001} (\bibinfo {year} {2018})},\ \Eprint {https://arxiv.org/abs/1708.09389} {arXiv:1708.09389 [hep-ph]} \BibitemShut {NoStop}%
\bibitem [{\citenamefont {Ahdida}\ \emph {et~al.}(2019)\citenamefont {Ahdida} \emph {et~al.}}]{SHiP:2018xqw}%
  \BibitemOpen
  \bibfield  {author} {\bibinfo {author} {\bibfnamefont {C.}~\bibnamefont {Ahdida}} \emph {et~al.} (\bibinfo {collaboration} {SHiP}),\ }\bibfield  {title} {\bibinfo {title} {{Sensitivity of the SHiP experiment to Heavy Neutral Leptons}},\ }\href {https://doi.org/10.1007/JHEP04(2019)077} {\bibfield  {journal} {\bibinfo  {journal} {JHEP}\ }\textbf {\bibinfo {volume} {04}},\ \bibinfo {pages} {077}},\ \Eprint {https://arxiv.org/abs/1811.00930} {arXiv:1811.00930 [hep-ph]} \BibitemShut {NoStop}%
\bibitem [{\citenamefont {Chou}\ \emph {et~al.}(2017)\citenamefont {Chou}, \citenamefont {Curtin},\ and\ \citenamefont {Lubatti}}]{Chou:2016lxi}%
  \BibitemOpen
  \bibfield  {author} {\bibinfo {author} {\bibfnamefont {J.~P.}\ \bibnamefont {Chou}}, \bibinfo {author} {\bibfnamefont {D.}~\bibnamefont {Curtin}},\ and\ \bibinfo {author} {\bibfnamefont {H.~J.}\ \bibnamefont {Lubatti}},\ }\bibfield  {title} {\bibinfo {title} {{New Detectors to Explore the Lifetime Frontier}},\ }\href {https://doi.org/10.1016/j.physletb.2017.01.043} {\bibfield  {journal} {\bibinfo  {journal} {Phys. Lett. B}\ }\textbf {\bibinfo {volume} {767}},\ \bibinfo {pages} {29} (\bibinfo {year} {2017})},\ \Eprint {https://arxiv.org/abs/1606.06298} {arXiv:1606.06298 [hep-ph]} \BibitemShut {NoStop}%
\bibitem [{\citenamefont {Aielli}\ \emph {et~al.}(2020)\citenamefont {Aielli} \emph {et~al.}}]{CODEX-b:2019jve}%
  \BibitemOpen
  \bibfield  {author} {\bibinfo {author} {\bibfnamefont {G.}~\bibnamefont {Aielli}} \emph {et~al.} (\bibinfo {collaboration} {CODEX-b}),\ }\bibfield  {title} {\bibinfo {title} {{Expression of interest for the CODEX-b detector}},\ }\href {https://doi.org/10.1140/epjc/s10052-020-08711-3} {\bibfield  {journal} {\bibinfo  {journal} {Eur. Phys. J. C}\ }\textbf {\bibinfo {volume} {80}},\ \bibinfo {pages} {1177} (\bibinfo {year} {2020})},\ \Eprint {https://arxiv.org/abs/1911.00481} {arXiv:1911.00481 [hep-ex]} \BibitemShut {NoStop}%
\bibitem [{\citenamefont {Hirsch}\ and\ \citenamefont {Wang}(2020)}]{Hirsch:2020klk}%
  \BibitemOpen
  \bibfield  {author} {\bibinfo {author} {\bibfnamefont {M.}~\bibnamefont {Hirsch}}\ and\ \bibinfo {author} {\bibfnamefont {Z.~S.}\ \bibnamefont {Wang}},\ }\bibfield  {title} {\bibinfo {title} {{Heavy neutral leptons at ANUBIS}},\ }\href {https://doi.org/10.1103/PhysRevD.101.055034} {\bibfield  {journal} {\bibinfo  {journal} {Phys. Rev. D}\ }\textbf {\bibinfo {volume} {101}},\ \bibinfo {pages} {055034} (\bibinfo {year} {2020})},\ \Eprint {https://arxiv.org/abs/2001.04750} {arXiv:2001.04750 [hep-ph]} \BibitemShut {NoStop}%
\bibitem [{\citenamefont {Chun}\ \emph {et~al.}(2019)\citenamefont {Chun}, \citenamefont {Das}, \citenamefont {Mandal}, \citenamefont {Mitra},\ and\ \citenamefont {Sinha}}]{Chun:2019nwi}%
  \BibitemOpen
  \bibfield  {author} {\bibinfo {author} {\bibfnamefont {E.~J.}\ \bibnamefont {Chun}}, \bibinfo {author} {\bibfnamefont {A.}~\bibnamefont {Das}}, \bibinfo {author} {\bibfnamefont {S.}~\bibnamefont {Mandal}}, \bibinfo {author} {\bibfnamefont {M.}~\bibnamefont {Mitra}},\ and\ \bibinfo {author} {\bibfnamefont {N.}~\bibnamefont {Sinha}},\ }\bibfield  {title} {\bibinfo {title} {{Sensitivity of Lepton Number Violating Meson Decays in Different Experiments}},\ }\href {https://doi.org/10.1103/PhysRevD.100.095022} {\bibfield  {journal} {\bibinfo  {journal} {Phys. Rev. D}\ }\textbf {\bibinfo {volume} {100}},\ \bibinfo {pages} {095022} (\bibinfo {year} {2019})},\ \Eprint {https://arxiv.org/abs/1908.09562} {arXiv:1908.09562 [hep-ph]} \BibitemShut {NoStop}%
\bibitem [{\citenamefont {Antusch}\ \emph {et~al.}(2017{\natexlab{a}})\citenamefont {Antusch}, \citenamefont {Cazzato},\ and\ \citenamefont {Fischer}}]{Antusch:2017hhu}%
  \BibitemOpen
  \bibfield  {author} {\bibinfo {author} {\bibfnamefont {S.}~\bibnamefont {Antusch}}, \bibinfo {author} {\bibfnamefont {E.}~\bibnamefont {Cazzato}},\ and\ \bibinfo {author} {\bibfnamefont {O.}~\bibnamefont {Fischer}},\ }\bibfield  {title} {\bibinfo {title} {{Sterile neutrino searches via displaced vertices at LHCb}},\ }\href {https://doi.org/10.1016/j.physletb.2017.09.057} {\bibfield  {journal} {\bibinfo  {journal} {Phys. Lett. B}\ }\textbf {\bibinfo {volume} {774}},\ \bibinfo {pages} {114} (\bibinfo {year} {2017}{\natexlab{a}})},\ \Eprint {https://arxiv.org/abs/1706.05990} {arXiv:1706.05990 [hep-ph]} \BibitemShut {NoStop}%
\bibitem [{\citenamefont {Drewes}\ and\ \citenamefont {Hajer}(2020)}]{Drewes:2019fou}%
  \BibitemOpen
  \bibfield  {author} {\bibinfo {author} {\bibfnamefont {M.}~\bibnamefont {Drewes}}\ and\ \bibinfo {author} {\bibfnamefont {J.}~\bibnamefont {Hajer}},\ }\bibfield  {title} {\bibinfo {title} {{Heavy Neutrinos in displaced vertex searches at the LHC and HL-LHC}},\ }\href {https://doi.org/10.1007/JHEP02(2020)070} {\bibfield  {journal} {\bibinfo  {journal} {JHEP}\ }\textbf {\bibinfo {volume} {02}},\ \bibinfo {pages} {070}},\ \Eprint {https://arxiv.org/abs/1903.06100} {arXiv:1903.06100 [hep-ph]} \BibitemShut {NoStop}%
\bibitem [{\citenamefont {Blondel}\ \emph {et~al.}(2022)\citenamefont {Blondel} \emph {et~al.}}]{Blondel:2022qqo}%
  \BibitemOpen
  \bibfield  {author} {\bibinfo {author} {\bibfnamefont {A.}~\bibnamefont {Blondel}} \emph {et~al.},\ }\bibfield  {title} {\bibinfo {title} {{Searches for long-lived particles at the future FCC-ee}},\ }\href {https://doi.org/10.3389/fphy.2022.967881} {\bibfield  {journal} {\bibinfo  {journal} {Front. in Phys.}\ }\textbf {\bibinfo {volume} {10}},\ \bibinfo {pages} {967881} (\bibinfo {year} {2022})},\ \Eprint {https://arxiv.org/abs/2203.05502} {arXiv:2203.05502 [hep-ex]} \BibitemShut {NoStop}%
\bibitem [{\citenamefont {Blondel}\ \emph {et~al.}(2016)\citenamefont {Blondel}, \citenamefont {Graverini}, \citenamefont {Serra},\ and\ \citenamefont {Shaposhnikov}}]{Blondel:2014bra}%
  \BibitemOpen
  \bibfield  {author} {\bibinfo {author} {\bibfnamefont {A.}~\bibnamefont {Blondel}}, \bibinfo {author} {\bibfnamefont {E.}~\bibnamefont {Graverini}}, \bibinfo {author} {\bibfnamefont {N.}~\bibnamefont {Serra}},\ and\ \bibinfo {author} {\bibfnamefont {M.}~\bibnamefont {Shaposhnikov}} (\bibinfo {collaboration} {FCC-ee study Team}),\ }\bibfield  {title} {\bibinfo {title} {{Search for Heavy Right Handed Neutrinos at the FCC-ee}},\ }\href {https://doi.org/10.1016/j.nuclphysbps.2015.09.304} {\bibfield  {journal} {\bibinfo  {journal} {Nucl. Part. Phys. Proc.}\ }\textbf {\bibinfo {volume} {273-275}},\ \bibinfo {pages} {1883} (\bibinfo {year} {2016})},\ \Eprint {https://arxiv.org/abs/1411.5230} {arXiv:1411.5230 [hep-ex]} \BibitemShut {NoStop}%
\bibitem [{\citenamefont {Antusch}\ \emph {et~al.}(2020)\citenamefont {Antusch}, \citenamefont {Fischer},\ and\ \citenamefont {Hammad}}]{Antusch:2019eiz}%
  \BibitemOpen
  \bibfield  {author} {\bibinfo {author} {\bibfnamefont {S.}~\bibnamefont {Antusch}}, \bibinfo {author} {\bibfnamefont {O.}~\bibnamefont {Fischer}},\ and\ \bibinfo {author} {\bibfnamefont {A.}~\bibnamefont {Hammad}},\ }\bibfield  {title} {\bibinfo {title} {{Lepton-Trijet and Displaced Vertex Searches for Heavy Neutrinos at Future Electron-Proton Colliders}},\ }\href {https://doi.org/10.1007/JHEP03(2020)110} {\bibfield  {journal} {\bibinfo  {journal} {JHEP}\ }\textbf {\bibinfo {volume} {03}},\ \bibinfo {pages} {110}},\ \Eprint {https://arxiv.org/abs/1908.02852} {arXiv:1908.02852 [hep-ph]} \BibitemShut {NoStop}%
\bibitem [{\citenamefont {Pascoli}\ \emph {et~al.}(2019)\citenamefont {Pascoli}, \citenamefont {Ruiz},\ and\ \citenamefont {Weiland}}]{Pascoli:2018heg}%
  \BibitemOpen
  \bibfield  {author} {\bibinfo {author} {\bibfnamefont {S.}~\bibnamefont {Pascoli}}, \bibinfo {author} {\bibfnamefont {R.}~\bibnamefont {Ruiz}},\ and\ \bibinfo {author} {\bibfnamefont {C.}~\bibnamefont {Weiland}},\ }\bibfield  {title} {\bibinfo {title} {{Heavy neutrinos with dynamic jet vetoes: multilepton searches at $ \sqrt{s}=14 $ , 27, and 100 TeV}},\ }\href {https://doi.org/10.1007/JHEP06(2019)049} {\bibfield  {journal} {\bibinfo  {journal} {JHEP}\ }\textbf {\bibinfo {volume} {06}},\ \bibinfo {pages} {049}},\ \Eprint {https://arxiv.org/abs/1812.08750} {arXiv:1812.08750 [hep-ph]} \BibitemShut {NoStop}%
\bibitem [{Beh(2013)}]{Behnke:2013xla}%
  \BibitemOpen
  \href@noop {} {\bibinfo {title} {{The International Linear Collider Technical Design Report - Volume 1: Executive Summary}}} (\bibinfo {year} {2013}),\ \Eprint {https://arxiv.org/abs/1306.6327} {arXiv:1306.6327 [physics.acc-ph]} \BibitemShut {NoStop}%
\bibitem [{\citenamefont {Antusch}\ \emph {et~al.}(2017{\natexlab{b}})\citenamefont {Antusch}, \citenamefont {Cazzato},\ and\ \citenamefont {Fischer}}]{Antusch:2016ejd}%
  \BibitemOpen
  \bibfield  {author} {\bibinfo {author} {\bibfnamefont {S.}~\bibnamefont {Antusch}}, \bibinfo {author} {\bibfnamefont {E.}~\bibnamefont {Cazzato}},\ and\ \bibinfo {author} {\bibfnamefont {O.}~\bibnamefont {Fischer}},\ }\bibfield  {title} {\bibinfo {title} {{Sterile neutrino searches at future $e^-e^+$, $pp$, and $e^-p$ colliders}},\ }\href {https://doi.org/10.1142/S0217751X17500786} {\bibfield  {journal} {\bibinfo  {journal} {Int. J. Mod. Phys. A}\ }\textbf {\bibinfo {volume} {32}},\ \bibinfo {pages} {1750078} (\bibinfo {year} {2017}{\natexlab{b}})},\ \Eprint {https://arxiv.org/abs/1612.02728} {arXiv:1612.02728 [hep-ph]} \BibitemShut {NoStop}%
\bibitem [{\citenamefont {Boyarsky}\ \emph {et~al.}(2023)\citenamefont {Boyarsky}, \citenamefont {Mikulenko}, \citenamefont {Ovchynnikov},\ and\ \citenamefont {Shchutska}}]{Boyarsky:2022epg}%
  \BibitemOpen
  \bibfield  {author} {\bibinfo {author} {\bibfnamefont {A.}~\bibnamefont {Boyarsky}}, \bibinfo {author} {\bibfnamefont {O.}~\bibnamefont {Mikulenko}}, \bibinfo {author} {\bibfnamefont {M.}~\bibnamefont {Ovchynnikov}},\ and\ \bibinfo {author} {\bibfnamefont {L.}~\bibnamefont {Shchutska}},\ }\bibfield  {title} {\bibinfo {title} {{Exploring the potential of FCC-hh to search for particles from B mesons}},\ }\href {https://doi.org/10.1007/JHEP01(2023)042} {\bibfield  {journal} {\bibinfo  {journal} {JHEP}\ }\textbf {\bibinfo {volume} {01}},\ \bibinfo {pages} {042}},\ \Eprint {https://arxiv.org/abs/2204.01622} {arXiv:2204.01622 [hep-ph]} \BibitemShut {NoStop}%
\bibitem [{\citenamefont {Li}\ \emph {et~al.}(2023)\citenamefont {Li}, \citenamefont {Liu},\ and\ \citenamefont {Lyu}}]{Li:2023tbx}%
  \BibitemOpen
  \bibfield  {author} {\bibinfo {author} {\bibfnamefont {P.}~\bibnamefont {Li}}, \bibinfo {author} {\bibfnamefont {Z.}~\bibnamefont {Liu}},\ and\ \bibinfo {author} {\bibfnamefont {K.-F.}\ \bibnamefont {Lyu}},\ }\bibfield  {title} {\bibinfo {title} {{Heavy neutral leptons at muon colliders}},\ }\href {https://doi.org/10.1007/JHEP03(2023)231} {\bibfield  {journal} {\bibinfo  {journal} {JHEP}\ }\textbf {\bibinfo {volume} {03}},\ \bibinfo {pages} {231}},\ \Eprint {https://arxiv.org/abs/2301.07117} {arXiv:2301.07117 [hep-ph]} \BibitemShut {NoStop}%
\bibitem [{\citenamefont {Kitano}\ \emph {et~al.}(2025)\citenamefont {Kitano}, \citenamefont {Low}, \citenamefont {Matsudo}, \citenamefont {Okawa},\ and\ \citenamefont {Roy}}]{Kitano:2025xaj}%
  \BibitemOpen
  \bibfield  {author} {\bibinfo {author} {\bibfnamefont {R.}~\bibnamefont {Kitano}}, \bibinfo {author} {\bibfnamefont {I.}~\bibnamefont {Low}}, \bibinfo {author} {\bibfnamefont {R.}~\bibnamefont {Matsudo}}, \bibinfo {author} {\bibfnamefont {S.}~\bibnamefont {Okawa}},\ and\ \bibinfo {author} {\bibfnamefont {S.}~\bibnamefont {Roy}},\ }\href@noop {} {\bibinfo {title} {{Heavy Neutral Lepton at Same-Sign Muon Collider}}} (\bibinfo {year} {2025}),\ \Eprint {https://arxiv.org/abs/2510.18390} {arXiv:2510.18390 [hep-ph]} \BibitemShut {NoStop}%
\bibitem [{\citenamefont {Bolton}\ \emph {et~al.}(2020)\citenamefont {Bolton}, \citenamefont {Deppisch},\ and\ \citenamefont {Bhupal~Dev}}]{Bolton:2019pcu}%
  \BibitemOpen
  \bibfield  {author} {\bibinfo {author} {\bibfnamefont {P.~D.}\ \bibnamefont {Bolton}}, \bibinfo {author} {\bibfnamefont {F.~F.}\ \bibnamefont {Deppisch}},\ and\ \bibinfo {author} {\bibfnamefont {P.~S.}\ \bibnamefont {Bhupal~Dev}},\ }\bibfield  {title} {\bibinfo {title} {{Neutrinoless double beta decay versus other probes of heavy sterile neutrinos}},\ }\href {https://doi.org/10.1007/JHEP03(2020)170} {\bibfield  {journal} {\bibinfo  {journal} {JHEP}\ }\textbf {\bibinfo {volume} {03}},\ \bibinfo {pages} {170}},\ \Eprint {https://arxiv.org/abs/1912.03058} {arXiv:1912.03058 [hep-ph]} \BibitemShut {NoStop}%
\bibitem [{\citenamefont {Fern{\'a}ndez-Mart{\'\i}nez}\ \emph {et~al.}(2023)\citenamefont {Fern{\'a}ndez-Mart{\'\i}nez}, \citenamefont {Gonz{\'a}lez-L{\'o}pez}, \citenamefont {Hern{\'a}ndez-Garc{\'\i}a}, \citenamefont {Hostert},\ and\ \citenamefont {L{\'o}pez-Pav{\'o}n}}]{Fernandez-Martinez:2023phj}%
  \BibitemOpen
  \bibfield  {author} {\bibinfo {author} {\bibfnamefont {E.}~\bibnamefont {Fern{\'a}ndez-Mart{\'\i}nez}}, \bibinfo {author} {\bibfnamefont {M.}~\bibnamefont {Gonz{\'a}lez-L{\'o}pez}}, \bibinfo {author} {\bibfnamefont {J.}~\bibnamefont {Hern{\'a}ndez-Garc{\'\i}a}}, \bibinfo {author} {\bibfnamefont {M.}~\bibnamefont {Hostert}},\ and\ \bibinfo {author} {\bibfnamefont {J.}~\bibnamefont {L{\'o}pez-Pav{\'o}n}},\ }\bibfield  {title} {\bibinfo {title} {{Effective portals to heavy neutral leptons}},\ }\href {https://doi.org/10.1007/JHEP09(2023)001} {\bibfield  {journal} {\bibinfo  {journal} {JHEP}\ }\textbf {\bibinfo {volume} {09}},\ \bibinfo {pages} {001}},\ \Eprint {https://arxiv.org/abs/2304.06772} {arXiv:2304.06772 [hep-ph]} \BibitemShut {NoStop}%
\bibitem [{\citenamefont {Vincent}\ \emph {et~al.}(2015)\citenamefont {Vincent}, \citenamefont {Martinez}, \citenamefont {Hern{\'a}ndez}, \citenamefont {Lattanzi},\ and\ \citenamefont {Mena}}]{Vincent:2014rja}%
  \BibitemOpen
  \bibfield  {author} {\bibinfo {author} {\bibfnamefont {A.~C.}\ \bibnamefont {Vincent}}, \bibinfo {author} {\bibfnamefont {E.~F.}\ \bibnamefont {Martinez}}, \bibinfo {author} {\bibfnamefont {P.}~\bibnamefont {Hern{\'a}ndez}}, \bibinfo {author} {\bibfnamefont {M.}~\bibnamefont {Lattanzi}},\ and\ \bibinfo {author} {\bibfnamefont {O.}~\bibnamefont {Mena}},\ }\bibfield  {title} {\bibinfo {title} {{Revisiting cosmological bounds on sterile neutrinos}},\ }\href {https://doi.org/10.1088/1475-7516/2015/04/006} {\bibfield  {journal} {\bibinfo  {journal} {JCAP}\ }\textbf {\bibinfo {volume} {04}},\ \bibinfo {pages} {006}},\ \Eprint {https://arxiv.org/abs/1408.1956} {arXiv:1408.1956 [astro-ph.CO]} \BibitemShut {NoStop}%
\bibitem [{\citenamefont {Langhoff}\ \emph {et~al.}(2022)\citenamefont {Langhoff}, \citenamefont {Outmezguine},\ and\ \citenamefont {Rodd}}]{Langhoff:2022bij}%
  \BibitemOpen
  \bibfield  {author} {\bibinfo {author} {\bibfnamefont {K.}~\bibnamefont {Langhoff}}, \bibinfo {author} {\bibfnamefont {N.~J.}\ \bibnamefont {Outmezguine}},\ and\ \bibinfo {author} {\bibfnamefont {N.~L.}\ \bibnamefont {Rodd}},\ }\bibfield  {title} {\bibinfo {title} {{Irreducible Axion Background}},\ }\href {https://doi.org/10.1103/PhysRevLett.129.241101} {\bibfield  {journal} {\bibinfo  {journal} {Phys. Rev. Lett.}\ }\textbf {\bibinfo {volume} {129}},\ \bibinfo {pages} {241101} (\bibinfo {year} {2022})},\ \Eprint {https://arxiv.org/abs/2209.06216} {arXiv:2209.06216 [hep-ph]} \BibitemShut {NoStop}%
\bibitem [{\citenamefont {Hernandez}\ \emph {et~al.}(2014)\citenamefont {Hernandez}, \citenamefont {Kekic},\ and\ \citenamefont {Lopez-Pavon}}]{Hernandez:2014fha}%
  \BibitemOpen
  \bibfield  {author} {\bibinfo {author} {\bibfnamefont {P.}~\bibnamefont {Hernandez}}, \bibinfo {author} {\bibfnamefont {M.}~\bibnamefont {Kekic}},\ and\ \bibinfo {author} {\bibfnamefont {J.}~\bibnamefont {Lopez-Pavon}},\ }\bibfield  {title} {\bibinfo {title} {{$N_{\rm eff}$ in low-scale seesaw models versus the lightest neutrino mass}},\ }\href {https://doi.org/10.1103/PhysRevD.90.065033} {\bibfield  {journal} {\bibinfo  {journal} {Phys. Rev. D}\ }\textbf {\bibinfo {volume} {90}},\ \bibinfo {pages} {065033} (\bibinfo {year} {2014})},\ \Eprint {https://arxiv.org/abs/1406.2961} {arXiv:1406.2961 [hep-ph]} \BibitemShut {NoStop}%
\bibitem [{\citenamefont {Dolgov}\ \emph {et~al.}(2000)\citenamefont {Dolgov}, \citenamefont {Hansen}, \citenamefont {Raffelt},\ and\ \citenamefont {Semikoz}}]{Dolgov:2000pj}%
  \BibitemOpen
  \bibfield  {author} {\bibinfo {author} {\bibfnamefont {A.~D.}\ \bibnamefont {Dolgov}}, \bibinfo {author} {\bibfnamefont {S.~H.}\ \bibnamefont {Hansen}}, \bibinfo {author} {\bibfnamefont {G.}~\bibnamefont {Raffelt}},\ and\ \bibinfo {author} {\bibfnamefont {D.~V.}\ \bibnamefont {Semikoz}},\ }\bibfield  {title} {\bibinfo {title} {{Cosmological and astrophysical bounds on a heavy sterile neutrino and the KARMEN anomaly}},\ }\href {https://doi.org/10.1016/S0550-3213(00)00203-0} {\bibfield  {journal} {\bibinfo  {journal} {Nucl. Phys. B}\ }\textbf {\bibinfo {volume} {580}},\ \bibinfo {pages} {331} (\bibinfo {year} {2000})},\ \Eprint {https://arxiv.org/abs/hep-ph/0002223} {arXiv:hep-ph/0002223} \BibitemShut {NoStop}%
\bibitem [{\citenamefont {Boyarsky}\ \emph {et~al.}(2009)\citenamefont {Boyarsky}, \citenamefont {Ruchayskiy},\ and\ \citenamefont {Shaposhnikov}}]{Boyarsky:2009ix}%
  \BibitemOpen
  \bibfield  {author} {\bibinfo {author} {\bibfnamefont {A.}~\bibnamefont {Boyarsky}}, \bibinfo {author} {\bibfnamefont {O.}~\bibnamefont {Ruchayskiy}},\ and\ \bibinfo {author} {\bibfnamefont {M.}~\bibnamefont {Shaposhnikov}},\ }\bibfield  {title} {\bibinfo {title} {{The Role of sterile neutrinos in cosmology and astrophysics}},\ }\href {https://doi.org/10.1146/annurev.nucl.010909.083654} {\bibfield  {journal} {\bibinfo  {journal} {Ann. Rev. Nucl. Part. Sci.}\ }\textbf {\bibinfo {volume} {59}},\ \bibinfo {pages} {191} (\bibinfo {year} {2009})},\ \Eprint {https://arxiv.org/abs/0901.0011} {arXiv:0901.0011 [hep-ph]} \BibitemShut {NoStop}%
\bibitem [{\citenamefont {Ruchayskiy}\ and\ \citenamefont {Ivashko}(2012)}]{Ruchayskiy:2012si}%
  \BibitemOpen
  \bibfield  {author} {\bibinfo {author} {\bibfnamefont {O.}~\bibnamefont {Ruchayskiy}}\ and\ \bibinfo {author} {\bibfnamefont {A.}~\bibnamefont {Ivashko}},\ }\bibfield  {title} {\bibinfo {title} {{Restrictions on the lifetime of sterile neutrinos from primordial nucleosynthesis}},\ }\href {https://doi.org/10.1088/1475-7516/2012/10/014} {\bibfield  {journal} {\bibinfo  {journal} {JCAP}\ }\textbf {\bibinfo {volume} {10}},\ \bibinfo {pages} {014}},\ \Eprint {https://arxiv.org/abs/1202.2841} {arXiv:1202.2841 [hep-ph]} \BibitemShut {NoStop}%
\bibitem [{\citenamefont {Gelmini}\ \emph {et~al.}(2020)\citenamefont {Gelmini}, \citenamefont {Kawasaki}, \citenamefont {Kusenko}, \citenamefont {Murai},\ and\ \citenamefont {Takhistov}}]{Gelmini:2020ekg}%
  \BibitemOpen
  \bibfield  {author} {\bibinfo {author} {\bibfnamefont {G.~B.}\ \bibnamefont {Gelmini}}, \bibinfo {author} {\bibfnamefont {M.}~\bibnamefont {Kawasaki}}, \bibinfo {author} {\bibfnamefont {A.}~\bibnamefont {Kusenko}}, \bibinfo {author} {\bibfnamefont {K.}~\bibnamefont {Murai}},\ and\ \bibinfo {author} {\bibfnamefont {V.}~\bibnamefont {Takhistov}},\ }\bibfield  {title} {\bibinfo {title} {{Big Bang Nucleosynthesis constraints on sterile neutrino and lepton asymmetry of the Universe}},\ }\href {https://doi.org/10.1088/1475-7516/2020/09/051} {\bibfield  {journal} {\bibinfo  {journal} {JCAP}\ }\textbf {\bibinfo {volume} {09}},\ \bibinfo {pages} {051}},\ \Eprint {https://arxiv.org/abs/2005.06721} {arXiv:2005.06721 [hep-ph]} \BibitemShut {NoStop}%
\bibitem [{\citenamefont {Sabti}\ \emph {et~al.}(2020)\citenamefont {Sabti}, \citenamefont {Magalich},\ and\ \citenamefont {Filimonova}}]{Sabti:2020yrt}%
  \BibitemOpen
  \bibfield  {author} {\bibinfo {author} {\bibfnamefont {N.}~\bibnamefont {Sabti}}, \bibinfo {author} {\bibfnamefont {A.}~\bibnamefont {Magalich}},\ and\ \bibinfo {author} {\bibfnamefont {A.}~\bibnamefont {Filimonova}},\ }\bibfield  {title} {\bibinfo {title} {{An Extended Analysis of Heavy Neutral Leptons during Big Bang Nucleosynthesis}},\ }\href {https://doi.org/10.1088/1475-7516/2020/11/056} {\bibfield  {journal} {\bibinfo  {journal} {JCAP}\ }\textbf {\bibinfo {volume} {11}},\ \bibinfo {pages} {056}},\ \Eprint {https://arxiv.org/abs/2006.07387} {arXiv:2006.07387 [hep-ph]} \BibitemShut {NoStop}%
\bibitem [{\citenamefont {Boyarsky}\ \emph {et~al.}(2021)\citenamefont {Boyarsky}, \citenamefont {Ovchynnikov}, \citenamefont {Ruchayskiy},\ and\ \citenamefont {Syvolap}}]{Boyarsky:2020dzc}%
  \BibitemOpen
  \bibfield  {author} {\bibinfo {author} {\bibfnamefont {A.}~\bibnamefont {Boyarsky}}, \bibinfo {author} {\bibfnamefont {M.}~\bibnamefont {Ovchynnikov}}, \bibinfo {author} {\bibfnamefont {O.}~\bibnamefont {Ruchayskiy}},\ and\ \bibinfo {author} {\bibfnamefont {V.}~\bibnamefont {Syvolap}},\ }\bibfield  {title} {\bibinfo {title} {{Improved big bang nucleosynthesis constraints on heavy neutral leptons}},\ }\href {https://doi.org/10.1103/PhysRevD.104.023517} {\bibfield  {journal} {\bibinfo  {journal} {Phys. Rev. D}\ }\textbf {\bibinfo {volume} {104}},\ \bibinfo {pages} {023517} (\bibinfo {year} {2021})},\ \Eprint {https://arxiv.org/abs/2008.00749} {arXiv:2008.00749 [hep-ph]} \BibitemShut {NoStop}%
\end{thebibliography}%
\appendix
\section{END MATTER}
\section{Accidental symmetry of the texture-zero solution}
The texture-zero solution discussed in the main text admits a simple symmetry interpretation. The exact vanishing of the light-neutrino mass originates from an emergent accidental global $U(1)$ symmetry associated with the zero mode of the renormalizable neutrino mass matrix. This symmetry is not imposed in the underlying theory, but emerges as a consequence of the texture-zero structure. The higher-dimensional operators introduced in Eq.~(\ref{eq:FN}), while preserving the underlying $U(1)_X$ symmetry, explicitly break this accidental symmetry, thereby lifting the protected zero mode and generating the light-neutrino mass.

The renormalizable neutrino mass matrix in Eq.~(\ref{eq:3x3_matrix}) possesses an exact
massless eigenstate satisfying
$
M_0|\nu\rangle=0$,
with normalized eigenvector
\begin{equation}
|\nu\rangle=
\frac{1}{\rho}
\left(
M_{12},
-m_2,
0
\right),
\qquad
\rho^2=m_2^2+M_{12}^2  \, .
\end{equation}
We define the projector onto the exact zero mode,
\begin{equation}
T\equiv|\nu\rangle\langle\nu|
=
\frac{1}{\rho^2}
\begin{pmatrix}
M_{12}^2 & -M_{12}m_2 & 0\\
-M_{12}m_2 & m_2^2 & 0\\
0 & 0 & 0
\end{pmatrix},
\end{equation}
which satisfies
$
TM_0=M_0T=0.
$
Since $T^2=T$, the transformation
\begin{equation}
U(\alpha)\equiv e^{i\alpha T}
=
1
+
\left(e^{i\alpha}-1\right)T
\end{equation}
leaves the neutrino mass matrix invariant for all~$\alpha$:
\begin{align}
U^T(\alpha)M_0U(\alpha)
&=
M_0
+
\left(e^{i\alpha}-1\right)TM_0
+
\left(e^{i\alpha}-1\right)M_0T
\nonumber\\
&
+
\left(e^{i\alpha}-1\right)^2TM_0T \, \, \,= \, \, \,
M_0 \, .
\end{align}
Therefore, the renormalizable neutrino mass matrix possesses an accidental global
$U(1)$ symmetry acting on the exact massless eigenstate,
\begin{equation}
U(1)_{\rm acc}:\qquad
|\nu\rangle
\rightarrow
e^{i\alpha}
|\nu\rangle \, .
\end{equation}
The higher-dimensional operators, defined in Eq.~(\ref{eq:FN}), modify the neutrino mass matrix according to
\begin{equation}
M=M_0+\Delta M,\qquad \Delta M=\begin{pmatrix}0&\mu_1&0\\\mu_1&\mu_2&0\\0&0&0\end{pmatrix} \, ,
\end{equation}
which satisfies
\begin{equation}
T\Delta M+\Delta MT\neq0 \,  ,
\end{equation}
demonstrating that the accidental symmetry is explicitly broken. The resulting lifting of the exact zero mode generates the
light-neutrino mass,
\begin{equation}
m_\nu
=
\langle\nu|\Delta M|\nu\rangle
=
{\rm Tr}(T\Delta M)
=
-2\mu_1
\frac{m_2M_{12}}{\rho^2}
+
\mu_2
\frac{m_2^2}{\rho^2} \, ,
\end{equation}
which reproduces Eq.~(\ref{mnu}). It is important to realize that
the higher-dimensional operators are invariant under the underlying
$U(1)_X$ symmetry, yet they constitute the leading explicit breaking of
the emergent $U(1)_{\rm acc}$ symmetry. Since these operators first
appear only at higher dimensions, the lifting of the zero mode is
naturally controlled by the effective-field-theory expansion parameter
$\epsilon=\langle\phi\rangle/\Lambda$. Consequently, the induced
light-neutrino mass inherits the same suppression.

We emphasize that the accidental symmetry
$U(1)_{\rm acc}$ is distinct from the underlying
$U(1)_X$ symmetry introduced in the ultraviolet theory.
The generator
\begin{equation}
T_X=\mathrm{diag}(q_{\nu_L},q_{N_1},q_{N_2})
      =\mathrm{diag}(0,1,0) \, ,
\end{equation}
which does not leave the texture-zero mass matrix invariant,
\begin{equation}
T_XM_0+M_0T_X=\begin{pmatrix}0&0&0\\0&0&M_{12}\\0&M_{12}&0\end{pmatrix}\neq0 \ ,
\end{equation}
whereas the projector
$T=|\nu\rangle\langle\nu|$
does.
Hence,
$U(1)_{\rm acc}$ is an emergent symmetry of the texture-zero
mass matrix associated with its exact zero mode, whereas
$U(1)_X$ is the underlying symmetry imposed on the ultraviolet
theory that determines the allowed operator structure. Although the
higher-dimensional operators preserve $U(1)_X$ by construction,
they provide the leading explicit breaking of the emergent
$U(1)_{\rm acc}$ symmetry.
\section{Additional Constraints on HNLs}
\label{app:HNLscons}
This section summarizes additional laboratory, astrophysical, and cosmological constraints on HNLs that complement the collider phenomenology discussed in the main text. Together, these constraints define the excluded and projected sensitivity regions shown in Fig.~\ref{fig:Const2}.

Complementary to the collider searches discussed in the main text, HNLs are also probed by rare meson ($A$) and $\tau$ decays,
such as
$A_1^\mp\rightarrow\ell_1^\pm\ell_2^\pm A_2^\mp$
and
$\tau^\pm\rightarrow\ell^\mp A_1^+A_2^-$,
whose branching fractions scale as
\begin{equation}
{\rm BR}
(A_1^\mp\rightarrow
\ell_1^\pm
\ell_2^\pm
A_2^\mp)
\sim
C
|V_{A_1}V_{A_2}|^2
|V_{\ell_1N}V_{\ell_2N}|,
\end{equation}
with
$C\simeq10^{-3},1,10^{-4}$ and
$10^{-5}$
for
$\tau$,
$K$,
$D_d/B_d$
and
$D_s$,
respectively.
The absence of such decays in present experiments places stringent limits on the
active--sterile mixing over the MeV--GeV mass range.
Although the present analysis focuses on the muon-flavor benchmark, the
underlying texture-zero mechanism is flavor independent and can be readily
generalized to all three lepton flavors, thereby opening the possibility of
rich flavor-dependent phenomenology.
The electron flavor channel receives additional constraints from neutrinoless
double-beta decay and pion/kaon peak searches,
while the tau flavor is comparatively less constrained and therefore offers an
interesting target for future dedicated searches.

The current experimental constraints, shown in grey shaded regions, from beam-dump, meson-decay, EW precision, and collider searches, such as  PIENU, PSI, KEK, T2K, LSND, BEBC, NA3, NuTeV, CHARM-II, DELPHI, LHC, NA62, $\mu$BooNE, and PMNS unitarity~\cite{Daum:1987bg, PIENU:2019usb, Bernardi:1985ny, Bernardi:1987ek, T2K:2019jwa,  Kelly:2021xbv, Arguelles:2021dqn, Hayano:1982wu, Yamazaki:1984sj, BNL-E949:2009dza, NA62:2021bji, WA66:1985mfx, NuTeV:1999kej, CHARM:1985nku, WA66:1985mfx, ATLAS:2019kpx, ATLAS:2022atq, CMS:2022fut, DELPHI:1996qcc, Fernandez-Martinez:2016lgt, Abdullahi:2022jlv, ATLAS:2024rzi, delAguila:2008pw, deBlas:2013gla, Antusch:2014woa, Bryman:2021teu, Blennow:2023mqx}, are summarized in Fig.~\ref{fig:Const2} in the $M_N-|V_{\mu N}|^2$ plane. Furthermore, the projected sensitivities of currently running experiments, such as DUNE, SHiP, and the HL-LHC (ATLAS, CMS, and LHCb), together with several proposed facilities, like FASER2, MATHUSLA, ANUBIS, FCC-ee, FCC-hh, FCC-he, ILC, and multi-TeV muon collider are displayed in various colored dashed lines~\cite{Krasnov:2019kdc, Ballett:2019bgd, Carbajal:2022zlp, Feng:2017uoz, SHiP:2018xqw, Chou:2016lxi, CODEX-b:2019jve, Hirsch:2020klk, Chun:2019nwi, Antusch:2017hhu, Drewes:2019fou, Blondel:2022qqo, Blondel:2014bra, Antusch:2019eiz, Pascoli:2018heg, Behnke:2013xla, Antusch:2016ejd, Boyarsky:2022epg, Li:2023tbx, Kitano:2025xaj} (for more details see Refs.~\cite{Bolton:2019pcu, Abdullahi:2022jlv, Fernandez-Martinez:2023phj} and references therein).

Besides direct laboratory searches, HNLs are also subject to
indirect cosmological and astrophysical constraints. Depending on their masses,
lifetimes, production history, and active--sterile mixings, HNLs may affect the
thermal evolution of the early Universe, leading to constraints from Big-Bang nucleosynthesis (BBN), the cosmic microwave background,
the effective number of relativistic degrees of freedom, and structure
formation and supernova cooling~\cite{Vincent:2014rja, Langhoff:2022bij, Hernandez:2014fha, Dolgov:2000pj, Boyarsky:2009ix,  Ruchayskiy:2012si, Gelmini:2020ekg, Sabti:2020yrt, Boyarsky:2020dzc}.
Cosmological observations may constrain the active--sterile mixing from both
below and above. For very small mixings the HNL lifetime can become
cosmologically long, whereas for intermediate mixings HNL decays around or
after the BBN epoch may modify the primordial light-element abundances or
inject entropy into the thermal plasma. For sufficiently large mixings, the
HNL lifetime becomes shorter than the onset of BBN, substantially relaxing the
corresponding cosmological bounds.
The corresponding region of interest is shown as the hatched area of Fig.~\ref{fig:Const2}, bounded by the dashed gray line.
Since these constraints depend on the
cosmological history and HNL production mechanism, they should be regarded as
indirect and complementary to the direct laboratory probes discussed above.
\end{document}